\documentclass[12pt]{article}

\usepackage{amsmath,amssymb,graphicx,multirow,xspace}
\usepackage[colorlinks=true,urlcolor=blue,anchorcolor=blue,citecolor=blue,filecolor=blue,linkcolor=blue,menucolor=blue,pagecolor=blue]{hyperref}
\usepackage[compress,numbers]{natbib}
\usepackage{placeins}
\usepackage{booktabs}
\usepackage{blindtext, rotating}
\usepackage{afterpage}
\usepackage{enumitem}
\usepackage{ marvosym }
\usepackage{verbatim}
\usepackage{authblk}
\usepackage{booktabs,multirow}

\usepackage{amsfonts}
\usepackage{float}
\usepackage{colortbl}
\usepackage{ccaption}
\usepackage{todonotes}
\usepackage{cancel}
\usepackage{verbatim}
\usepackage{color}
\usepackage{caption}
\usepackage{subcaption}
\usepackage{placeins}
\usepackage{fancyhdr}
\usepackage{epstopdf}
\usepackage{arydshln}
\usepackage{pifont}
\usepackage{slashed}
\usepackage{psfrag}
\usepackage{epstopdf}
\usepackage{multicol}
\usepackage{enumerate}
\usepackage[top=1in,bottom=1in,left=1in,right=1in,footskip=0.5in,headheight=0.5in,footnotesep=0.2in,marginparwidth=0in,marginparsep=0in]{geometry}
\usepackage{titlesec}

\newcommand{\abs}[1]{\big | #1 \big |}
\newcommand\fverb{\setbox\fverbbox=\hbox\bgroup\verb}
\newcommand\fverbdo{\egroup\medskip\noindent%
            \fbox{\unhbox\fverbbox}\ }
\newcommand\fverbit{\egroup\item[\fbox{\unhbox\fverbbox}]}
\newbox\fverbbox

\long\def\symbolfootnote[#1]#2{\begingroup%
\def\thefootnote{\fnsymbol{footnote}}\footnote[#1]{#2}\endgroup}

\newcommand{\be}{\begin{equation}}
\newcommand{\ee}{\end{equation}}

\newcommand{\mat}{\begin{pmatrix}}
\newcommand{\rix}{\end{pmatrix}}

\renewcommand{\bar}{\overline}

\newcommand{\qq}{\qquad}
\newcommand{\qqq}{\qquad\quad}
\newcommand{\qqqq}{\qquad\qquad}

\newcommand{\beqa}{\begin{eqnarray}}
\newcommand{\eeqa}{\end{eqnarray}}

\newcommand{\beq}{\begin{equation}}
\newcommand{\eeq}{\end{equation}}

\renewcommand{\arraystretch}{1.5}

\makeatletter
\DeclareRobustCommand{\cev}[1]{%
  \mathpalette\do@cev{#1}%
}
\newcommand{\do@cev}[2]{%
  \fix@cev{#1}{+}%
  \reflectbox{$\m@th#1\vec{\reflectbox{$\fix@cev{#1}{-}\m@th#1#2\fix@cev{#1}{+}$}}$}%
  \fix@cev{#1}{-}%
}
\newcommand{\fix@cev}[2]{%
  \ifx#1\displaystyle
    \mkern#23mu
  \else
    \ifx#1\textstyle
      \mkern#23mu
    \else
      \ifx#1\scriptstyle
        \mkern#22mu
      \else
        \mkern#22mu
      \fi
    \fi
  \fi
}

\makeatother

\begin{document}

\author{Daniel Egana-Ugrinovic}

\affil{ \small{C. N. Yang Institute for Theoretical Physics, Stony Brook, NY 11794}}
  \date{\vspace{-5ex}}
\title{The minimal fermionic model of electroweak baryogenesis}
\clearpage\maketitle
  \date{\vspace{-5ex}}
  \thispagestyle{empty}
 
 \begin{abstract}
We present the minimal model of electroweak baryogenesis induced by fermions. The model consists of an extension of the Standard Model with one electroweak singlet fermion and one pair of vector like doublet fermions with renormalizable couplings to the Higgs. A strong first order phase transition is radiatively induced by the singlet-doublet fermions, while the origin of the baryon asymmetry is due to asymmetric reflection of the same set of fermions on the expanding electroweak bubble wall. The singlet-doublet fermions are stabilized at the electroweak scale by chiral symmetries and the Higgs potential is stabilized by threshold corrections coming from a multi-TeV ultraviolet completion which does not play any significant role in the phase transition. We work in terms of background symmetry invariants and perform an analytic semiclassical calculation of the baryon asymmetry, showing that the model may effectively generate the observed baryon asymmetry for percent level values of the unique invariant CP violating phase of the singlet-doublet sector. We include a detailed study of electron electric dipole moment and electroweak precision limits, and  for one typical benchmark scenario we also recast existing collider constraints, showing that the model is consistent with all current experimental data. We point out that fermion induced electroweak baryogenesis has irreducible phenomenology at the $13 \, \textrm{TeV}$ LHC since the new fermions must be at the electroweak scale, have electroweak quantum numbers and couple strongly with the Higgs. The most promising searches involve topologies with multiple leptons and missing energy in the final state.
\end{abstract}
\newpage
\setcounter{page}{1}

\section{Introduction}

The explanation of the baryon asymmetry of the universe is one of the outstanding problems in particle physics. The only baryogenesis mechanism that we know of which necessarily requires new physics at the electroweak scale and is therefore most likely to be experimentally testable is electroweak baryogenesis (EWBG) \cite{Kuzmin:1985mm}, for reviews see  \cite{Cline:2006ts,Riotto:1998bt,Quiros:1999jp}. It relies on the nucleation of Higgs vacuum bubbles at the electroweak phase transition on which fermions reflect asymmetrically, creating an excess in some global charge which is processed into a baryon asymmetry by weak sphalerons \cite{Cohen:1990py,Nelson:1991ab}. For the mechanism to be effective, the Standard Model Higgs potential requires modifications in order to ensure the nucleation of bubbles with a Higgs condensate larger than the critical temperature of the phase transition. This is the strong first order phase transition requirement, which ensures that the baryon asymmetry is not washed out by the same weak sphalerons which create the asymmetry in the first place. For these bubbles to be nucleated at the critical temperature, an energy barrier in the effective potential is needed in order to separate the electroweak symmetric phase (outside the bubble) from the electroweak broken phase (inside the bubble). Also, a new source of CP violation is required for efficient generation of an asymmetry, since Standard Model CP violation is insufficient due to the suppression factors in the Jarlskog invariant \cite{Huet:1994jb}.

With the exception of \cite{Carena:2004ha,Fok:2008yg,Davoudiasl:2012tu,Fairbairn:2013xaa}, the literature has overwhelmingly concentrated in coupling new scalars to the Higgs in order to induce the strong first order phase transition, mostly because the barrier may be generated with a negative Higgs quartic stabilized by a threshold $(H^\dagger H)^3$ term as in \cite{Grojean:2004xa}, which at tree level may only be generated by integrating out heavy scalars, or because in a \textit{large temperature} expansion of the Higgs effective potential, scalars contribute to a negative cubic term which induces a barrier while fermions do not \cite{Cline:2006ts}. For a classification of the extensive literature on scalar models see \cite{Chung:2012vg}. However, new scalars introduce additional tuning in the theory, since they are not stable at the electroweak scale. Also, the simplest scalar models involve only singlets, in which case the only irreducible phenomenology involves  precision Higgs physics \cite{Noble:2007kk,Katz:2014bha,Curtin:2014jma,Huang:2016cjm} which may require a new high energy collider. 

This motivates us to revisit fermion induced electroweak baryogenesis. We draw additional motivation from the following observation. Both fermions and scalars contribute to the Higgs thermal potential if they obtain mass from the Higgs mechanism, and only in a large temperature expansion is clear that scalars contribute most efficiently to the barrier. If instead one performs a \textit{small temperature} expansion, one finds that the leading contribution to the thermal potential is \textit{exactly the same} for both fermions and bosons and is proportional to \cite{Anderson:1991zb}
\begin{equation}
 -\frac{T^2 m^2(\phi)}{2\pi^2} \, K_2\,\big(m(\phi)/T \big) + {\cal O}\big(\, T^2 m(\phi)^2 e^{-2m(\phi)/T} \, \big)
 \label{eq:thermalapprox}
\end{equation}
where $\phi$ is the Higgs field, $m(\phi)$  is the mass of the fermion or boson and $K_2$ a modified Bessel function. So in cases in which the critical temperature is smaller than the masses of the fermions contributing to the effective potential, fermionic models may be equally as effective as scalar models in inducing a barrier radiatively\footnote{Note however that the zero-temperature radiative effects are still different for fermions and scalars}. From \eqref{eq:thermalapprox}, we see that the key element is the the relation between the mass of the fermion and the Higgs condensate, so the problem reduces to identifying what type of mass relation leads to the formation of a barrier in the effective potential.

In the Standard Model, at temperatures right above the critical temperature for the phase transition, the Higgs effective potential around the origin of Higgs field space monotonically increases with the Higgs field, so there is no energy barrier leading to a strong first order phase transition \cite{Csikor:1998eu}. Introducing new chiral fermions at the electroweak scale which obtain their masses only from the Higgs condensate delays the phase transition \cite{Carena:2004ha}, but does not modify the picture around the origin of field space (even though at large field excursions new fermions lead to instabilities in the Higgs potential due to their zero temperature contributions). The reason is that the masses of chiral fermions and therefore their thermal potential \eqref{eq:thermalapprox} are monotonically increasing with the Higgs field, so the full Higgs effective potential retains the same qualitative behavior of the Standard Model effective potential around the origin of Higgs field space.

\begin{figure}[htbp]
\begin{center}
\includegraphics[width=10cm]{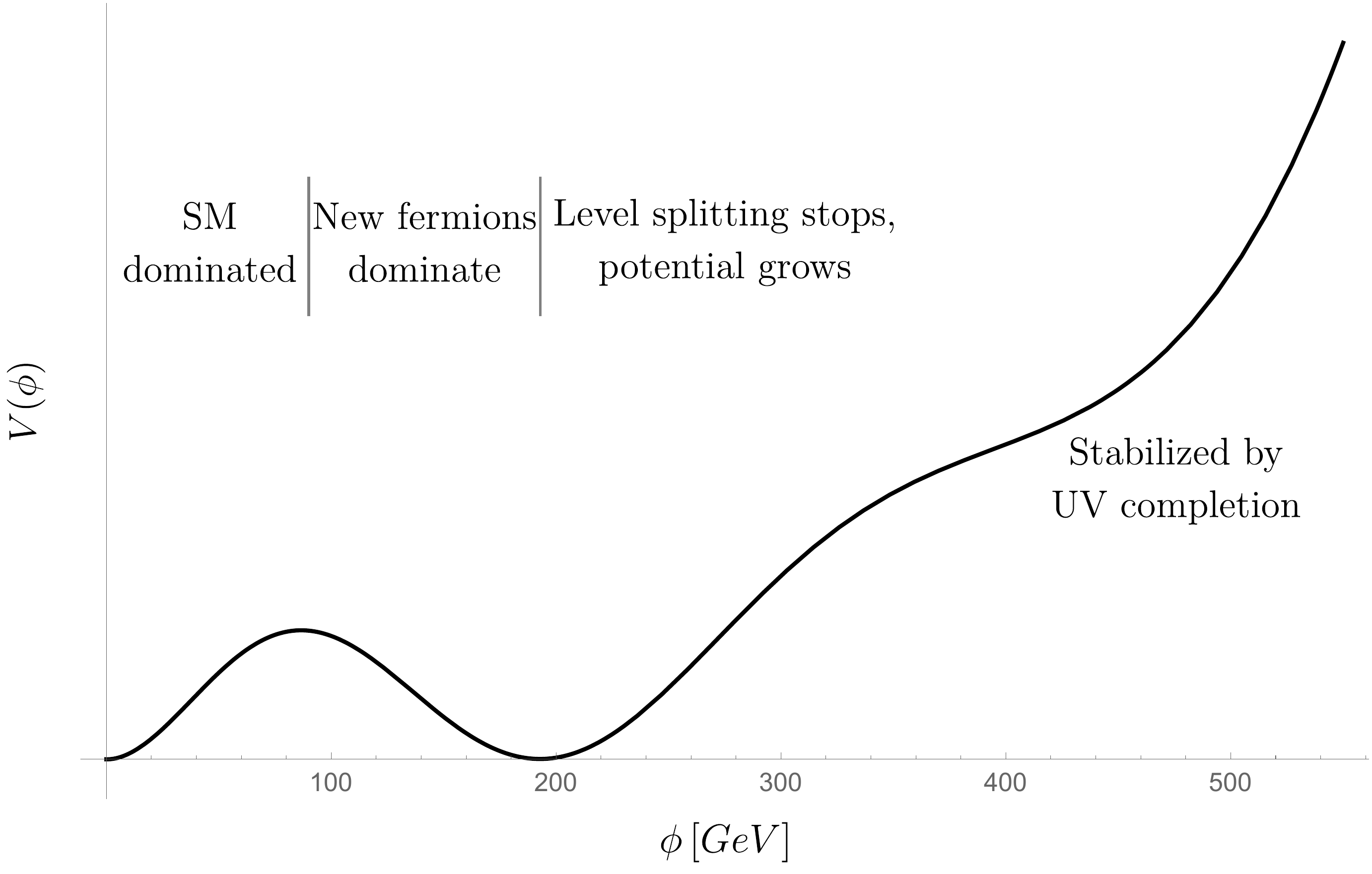}
\caption{Schematic representation of the Higgs effective potential as a function of the Higgs field $\phi$ at the critical temperature of the electroweak phase transition, in a model with a barrier induced by new electroweak-scale fermions.}
\label{fig:qualitative}
\end{center}
\end{figure}

The picture changes when we introduce new fermions that have both vector-like masses and masses obtained from mixing with other fermions in the electroweak broken vacuum. In this case, the masses of the fermions depend on the Higgs in a qualitatively different manner, since the condensate may induce level splitting, which reduces the mass of the lightest eigenstate of the mass matrix and increases the mass of the heavier ones. Schematically and around the origin of Higgs field space, the mass of the lightest new fermion is $m \sim M - y^2 \phi^2/M$, where $y$ is a renormalizable coupling between the new fermions and the Higgs, $M$ a vector like mass term and the second term represents level splitting. In this case, the mass of the lightest fermion \textit{decreases} with increasing values of the Higgs condensate $\phi$, leading to a reduction in the thermal effective potential. There is then a competition between the Standard Model terms (plus all polynomial counterterms), which tend to restore electroweak symmetry, and the new fermionic terms which have the opposite effect. In this work we present a simple model realizing the above mass relation, for which we find that there are large regions of parameter space in which at the critical temperature, around the origin the monotonically increasing Standard Model terms dominate while close the electroweak scale the negative contribution from the fermionic terms dominate. At field ranges $ \phi \sim M/{y}$ level splitting stops, the mass of the lightest fermion (and therefore its thermal potential) starts growing, and the potential is stabilized. Higher order Standard Model terms also help in stabilizing the potential. The summarized effect is the formation of an energy barrier separating the minimum at the origin of field space from a second minimum where electroweak symmetry is broken. At even larger field ranges and most importantly, at zero temperatures, the new fermions lead to an instability which the Standard Model thermal terms cannot counteract, so this minimal picture is insufficient. In order to solve this issue, we introduce stabilizing irrelevant operators of the type $(H^\dagger H)^n$ with $n \geq 3$, which may be interpreted as thresshold corrections coming from a multi-TeV UV completion \textit{which does not play any role} in the formation of the barrier (differently from \cite{Grojean:2004xa}), since the effects of the corresponding irrelevant operators at the electroweak scale are suppressed by powers of the electroweak scale over the TeV-scale cutoff of the UV completion. We present a schematic picture of the full mechanism in figure \ref{fig:qualitative}.

It is easy to find the minimal fermionic model leading to a strong first order phase transition by exhaustion. The two most minimal anomaly free extensions of the Standard Model with new fermions coupling to the Higgs at renormalizable level are, with one multiplet the right handed neutrino $\psi_S$ and with two multiplets a vector like doublet $\psi_L,\psi_{\bar{L}}$ \cite{Bizot:2015zaa}. In both cases, the new fermions couple to the Higgs by mixing with Standard Model fermions, and these couplings are generically strongly constrained \cite{Bizot:2015zaa}. Most importantly, neither the right handed neutrino model nor the vector like doublet model lead to level splitting, and they do not generate a strong first order phase transition. The next simplest fermionic extension of the Standard Model is a combination of the two models above and contains three fermion multiplets, one $SU(2)$ singlet $\psi_S$ \textit{and} a vector like $SU(2)$ doublet $\psi_L,\psi_{\bar{L}}$ \cite{Bizot:2015zaa,Mahbubani:2005pt,Cohen:2011ec,Abe:2014gua,Basirnia:2016szw,Calibbi:2015nha}. In this case, one can write down Yukawas with the Higgs without involving Standard Model fermions, $ \psi_{\bar{L}} \, H^c \,  \psi_S \,\,, \,\, \psi_{{L}} \, H \,  \psi_S$ \footnote{The gauge representation and hypercharges of the singlet-doublet fermions are entirely fixed by the requisite Yukawa couplings and anomaly cancellation.}. These Yukawas may be large, so these fermions may lead to a large effect on the Higgs effective potential at one loop. Moreover, in this singlet-doublet model the Higgs field induces level splitting between the neutral singlet-doublet fermions. 

In this work we demonstrate that the singlet-doublet model is in fact a complete realization of fermion induced electroweak baryogenesis, by showing that it leads to a strong first order phase transition, has the requisite CP violating phase leading to the generation of the baryon asymmetry and is consistent with all current experimental data. We also point out that the collider phenomenology of fermion induced electroweak baryogenesis is significantly different from the more popular models of scalar induced EWBG. Since in fermion induced EWBG the new fermions must be at the electroweak scale and have electroweak quantum numbers, they are pair produced and decay via electroweak gauge bosons and the Higgs leading to a rich set of final states, with the largest discovery potential in final states with multiple leptons and missing energy. 

To the best of our knowledge, the model presented here is the first complete implementation and phenomenological study of purely fermion induced EWBG. Previous works mostly follow the ideas of \cite{Carena:2004ha} and of \cite{Davoudiasl:2012tu}. In the seminal work \cite{Carena:2004ha} it was first realized that the strong first order phase transition may be induced by fermions in a supersymmetric context, but in that work the effective potential is radiatively stabilized by new scalars which lead to a contribution to the barrier, so it is not straightforward to quantify and study the effect of the fermions alone. Here we show with a simplified model that the barrier may be generated exclusively by fermions while simultaneously stabilizing the potential with threshold corrections without affecting the strength of the phase transition, we isolate the requirements for fermion induced EWBG to be effective and we identify the irreducible phenomenology. In \cite{Davoudiasl:2012tu} the barrier is generated by integrating out heavy fermions, but the baryon asymmetry is not explored and there is no proof that the potential may be stabilized without affecting the strength of the phase transition.

This paper is organized as follows. In section \ref{sec:model} we present the model. We carefully work throughout in terms of background symmetry invariants, in order to keep track of the unique CP violating phase of the model. In section \ref{sec:phasetransition} we numerically determine the strength of the phase transition from the full one-loop Higgs effective potential in the CP conserving case, we study electroweak precision limits and comment on the stability of the Higgs potential and Landau poles. In section \ref{sec:CPviolating} we include CP violation and study the corresponding phenomenology. We perform an analytic, semiclassical and background symmetry invariant calculation of the baryon asymmetry, and study electron electric dipole moment constraints. In section \ref{sec:results} we present and combine all the results, including the baryon asymmetry, strength of the phase transition, electroweak precision and electric dipole moment constraints. In section \ref{sec:collider} we briefly comment on the collider phenomenology. We conclude in section \ref{sec:conclusions}.

\section{Singlet-doublet model at finite temperatures}
\label{sec:model}
Consider the Standard Model extended with a fermionic singlet $\psi_S$ and a vector like electroweak doublet $\psi_L,\psi_{\bar{L}}$, with gauge charges defined in table \ref{tab:gaugecharges}. We assign a discrete ${\mathbb Z}_2$ charge to the singlet and doublet fermions, specified in table \ref{tab:gaugecharges}, which forbids mixing with the standard model fermions. The most general Lagrangian at the renormalizable level for the singlet-doublet fermions, the Higgs doublet with hypercharge $Y(H)=1$ and the SM fermions, respecting the discrete ${\mathbb Z}_2$ symmetry is
\begin{eqnarray}
D_\mu H^\dagger D^\mu H
+
i 
\psi_L ^{\dagger}\, \bar{\sigma}^\mu D_\mu \, \psi_L
+
i 
\psi_{\bar{L}} ^{\dagger}\, \bar{\sigma}^\mu D_\mu \, \psi_{\bar{L}}
+
i 
\psi_S ^{\dagger}\, \bar{\sigma}^\mu D_\mu \, \psi_S
\nonumber \\
  - 
  ~
  V(H) 
   - \bigg[
   ~  y^u_{ij} ~ Q_i H \bar{u}_j
   - y^d_{ij} Q_i H^c \bar{d}_j 
   - y^\ell_{\ij} L_i H^c \bar{\ell}_j
  \nonumber 
  \\
+
\frac{1}{2}m_S \, \psi_S \psi_S  
+
 m_L \psi_L \psi_{\bar{L}} 
-  \lambda_d \, \psi_{\bar{L}} \, H^c \,  \psi_S 
+  \lambda_u \, \psi_{L} H \, \psi_S 
+ \textrm{h.c.}
\,
\Big]
\label{eq:lagrangiansingletdoublet}
\end{eqnarray}
where the tree-level renormalizable Higgs potential is defined as
\begin{eqnarray}
V_\textrm{tree} 
\equiv
m^2 H^\dagger H 
+ \frac{\lambda}{2} (H^\dagger H)^2  
\label{eq:higgspontential1}
\end{eqnarray}
%
We normalize the Higgs condensate as 
\begin{equation}
{\phi^2 \over 2} \equiv \langle H^\dagger H \rangle  
\label{eq:vacuum}
\end{equation}
where without loss of generality we can work in a gauge with $\phi \geq 0$. At zero temperature the potential is minimized at $\phi(T=0)\equiv v = 246 \, \textrm{GeV}$.

\begin{table}[ht!]
\begin{center}
$$
\begin{array}{ccccc} 
& SU(3)_c & SU(2)_L & U(1)_Y  &  {\mathbb Z}_2
\\ \hline
\psi_S
&
{\mathbf 1}
&
{\mathbf 1}
&
0
&
-1
\\
\psi_L
&
{\mathbf 1}
&
{\mathbf 2}
&
-1
&
-1
\\
\psi_{\bar{L}}
&
{\mathbf 1}
&
{\mathbf 2}
&
1
&
-1
\\
\hline
\end{array}
$$
\end{center}
\caption{Field content of the singlet-doublet model. The singlet and doublet fermions are odd under the ${\mathbb Z}_2$, while the standard model fermions are even. The discrete symmetry forbids Yukawas involving the singlet and doublet fermions and standard model fermions and makes the lightest fermion of the singlet-doublet sector stable.}
\label{tab:gaugecharges}
\end{table}

The background (spurious) symmetry group of the model corresponds to Standard Model flavor group cross a $U(1)_S\times  U(1)_{L}  \times  U(1)_{\bar{L}}$ group specified in table \ref{tab:U1PQ2}. The singlet-doublet model contains five physical observables, or equivalently, five invariants under the $U(1)_S\times  U(1)_{L}  \times  U(1)_{\bar{L}}$ background symmetry. 

The CP even invariants are four, and may be chosen to be the absolute values of the singlet and doublet Lagrangian masses $\abs{m_S}, \abs{m_L}$ and the absolute values of the two Yukawa couplings $\abs{\lambda_u}, \abs{\lambda_d}$. In this work we will be interested in electroweak-scale values for the Lagrangian masses  $\abs{m_S}, \abs{m_L} \sim {\cal O} (10^2 \, {\textrm{GeV}})$, since heavier singlet-doublet fermions would decouple from the thermal plasma at the electroweak phase transition and would not lead to significant effects on the effective theory. This choice is of course technically natural: the smallness and stability of the singlet-doublet masses at the electroweak scale within any high scale UV completion is ensured by the chiral symmetries of the singlet-doublet sector\footnote{Choosing electroweak scale singlet-doublet Lagrangian masses leads to a coincidence of scales problem: in this theory there is no explicit (dynamical) relation between the singlet-doublet lagrangian masses and the electroweak scale itself. For brevity we will not comment any further on this problem, whose solution would require further details about the UV completion.}.

 The final remaining physical parameter of the theory is a unique CP odd invariant $\textrm{Im} \,  \lambda_u \lambda_d  m_S^* m_L^* $. In the case in which any of the parameters $\lambda_u,\lambda_d,m_S$ or $m_L$ are zero, the CP odd invariant vanishes and there is no effective CP violation in the theory. For non vanishing Yukawas and singlet-doublet masses the CP odd invariant may be traded for the invariant CP violating phase
\begin{equation}
\delta_{\textrm{CP}}\equiv \textrm{Arg}\big( \,  \lambda_u \lambda_d  m_S^* m_L^* \, \big) \in  0, 2\pi 
\label{eq:definitiondelta}
\end{equation} 
The singlet-doublet sector violates CP whenever $\delta_{\textrm{CP}} \neq 0 , \pi$ and conserves CP otherwise. $\delta_{\textrm{CP}}$ is the required source of CP violation for a baryon asymmetry to be obtained in this model. 


\begin{table}[ht!]
\begin{center}
$$
\begin{array}{ccccc} 
& U(1)_S & U(1)_L  &  U(1)_{\bar{L}} 
\\ \hline
\psi_S
&
-1
\\
\psi_L
&
&
-1
&
\\
\psi_{\bar{L}}
&
&
&
-1
\\
\hline
m_S 
&
2
&
&
\\
m_L 
&
&
1
&
1
\\
\lambda_u
&
1
&
1
&
\\
\lambda_d
&
1
&
&
1
\\
  \hline
\end{array}
$$
\end{center}
\caption{Background $U(1)_S\times  U(1)_{L}  \times  U(1)_{\bar{L}}$ charges of the Singlet-Doublet model. All the Standard Model fields are neutral under the background symmetry group $U(1)_S\times  U(1)_{L}  \times  U(1)_{\bar{L}}$, while the Singlet-Doublet fermions are neutral under the non-abelian Standard Model flavor group. Note that the CP violating phase $\delta_{\textrm{CP}}$ defined in equation \eqref{eq:definitiondelta} is a CP odd invariant.}
\label{tab:U1PQ2}
\end{table}
We define the charged an neutral components of the fermionic doublets as 
\begin{equation}
\psi_L \,\, \equiv  \,\,
{\def\arraystretch{1.2}\tabcolsep=10pt
\left(
\begin{array}{c}
   \psi_L^0  \\ 
     \psi_- \\ 
  \end{array}
       \right)
       }
\qq
\psi_{\bar L}
\,\,
\equiv
\,\,
{\def\arraystretch{1.2}\tabcolsep=10pt
\left(
 \begin{array}{c}
   \psi_+  \\ 
     \psi_{\bar{L}}^0 \\ 
  \end{array}
  \right)
  }
  \label{eq:componentdefinitions}
\end{equation}
The spectrum of the theory consists of one charged Dirac pair formed with $\psi_+$ and $\psi_-$ and three neutral Majorana fields. The charged fields $\psi_\pm$ do not couple to the Higgs, so their non-negative, background symmetry invariant Dirac mass squared is
\begin{equation}
(m_F^{\pm})^2= \abs{m_L}^2
\end{equation}
On the other hand, the symmetric complex mass matrix for the neutral Majorana fields $\psi_S, \psi_L^0, \psi_{\bar{L}}^0$ in the electroweak broken vacuum defined in \eqref{eq:vacuum} is 
\begin{equation}
{\mathcal M}
\equiv
\left(\begin{array}{ccc} m_S & \frac{\lambda_u \phi}{\sqrt{2}} & \frac{\lambda_d \phi}{\sqrt{2}} \\ \frac{\lambda_u \phi}{\sqrt{2}} & 0 & m_L \\   \frac{\lambda_d \phi}{\sqrt{2}} & m_L & 0\end{array}\right)
=
U
\,
\left(\begin{array}{ccc} 
m_1 & 0 & 0  \\  0 & m_2 & 0 \\   0 & 0 & m_3 \end{array}\right)
\,
U^T
\label{eq:Mf}
\end{equation}
where the mass singular values $m_i$, $i=1,2,3$ corresponding to the mass eigenstates $\psi_i$ are by definition non-negative and the  matrix $U$ is a unitary singular value decomposition matrix, which is defined by \eqref{eq:Mf} only up to a reparametrization symmetry independent of the background symmetry, corresponding to right multiplication by a discrete unitary matrix. Under a background symmetry transformation the matrix $U$ transforms by left multiplication with a diagonal unitary matrix with the charges specified in \eqref{tab:U1PQ2}, $\textrm{diag}(e^{-i\alpha}, e^{-i\beta},  e^{-i\gamma})$, where $\alpha, \beta$ and $\gamma$ are arbitrary phases. Physical observables are invariants under both the discrete reparametrization and background symmetry transformations. For instance, the three mass singular values are invariants. To make this explicit, note that the hermitian mass squared matrix is
\begin{equation}
{\mathcal M}^\dagger {\mathcal M}
=
\left(\begin{array}{ccc}
\abs{m_S}^2
+ 
\frac{\phi^2}{2}
\big[
\, \abs{\lambda_u}^2
+
\abs{\lambda_d}^2
\,
\big]
& 
\frac{1}{\sqrt{2}}
\big[
\,
\phi
\lambda_u 
m_S^*
+ 
\phi
m_L 
\lambda_d^*
\,
\big]
&
\frac{1}{\sqrt{2}}
\big[
\,
\phi 
m_S^*
\lambda_d 
+ 
\phi
m_L 
\lambda_u^*
\,
\big]
\\ 
& 
\abs{m_L}^2 
+ 
\frac{1}{2} 
\phi^2
\abs{\lambda_u}^2
& 
\frac{1}{2}
\phi^2
\lambda_d 
\lambda_u^*
\\
&  
& 
\abs{m_L}^2 
+ 
\frac{1}{2} 
\phi^2
\abs{\lambda_d}^2
\end{array}\right)
\label{eq:Mfsquared}
\end{equation}
which has a characteristic equation given by
\begin{equation}
  -\det({\cal M}^\dagger {\cal M} -xI)= x^3 + a x^2
  +bx
  +c=0  
  \label{eq:characeq}
\end{equation}
with coefficients
\begin{eqnarray}
  a\! \! \! &\equiv&\! \! \!   
  -2 
 \abs{m_L}^2
  - 
 \abs{m_S}^2
  -
  \big[
  \,
    \abs{\lambda_u}^2
    +
  \abs{\lambda_d}^2
  \,
  \big]
   \,\phi^2
   \nonumber \\
  b\! \! \! &\equiv&\! \! \! 
\abs{m_L}^4
  +  
  2
  \abs{m_L m_S}^2
  + \Big(
  \,
  \big[\, 
  \abs{\lambda_u}^2
  +
    \abs{\lambda_d}^2 
    \big]
    \,
  \, \frac{\phi^2}{2}
     \,
   \Big)^2
  \nonumber \\
  &+& 
  \abs{m_L}^2
  \big[
  \,
  \abs{\lambda_u}^2
  +
    \abs{\lambda_d}^2
    \,
    \big]
    \,\phi^2
    -
    \,
    \big[
    \,
       m_L^*
    m_S^*
    \lambda_u
    \lambda_d
    +
    \textrm{h.c.}
    \,
    \big]
     \,\phi^2
    \nonumber \\
  c\! \! \! &\equiv&\! \! \!   
  -\abs{m_L}^4 
  \abs{m_S}^2
  +
  \abs{m_L}^2
  \big[
  \,
    m_L^*
  m_S^*
  \lambda_u
  \lambda_d
  +
  \textrm{h.c.}
  \,
  \big]
  \,\phi^2
  -
  \abs{
  m_L
  \lambda_u
  \lambda_d
 \, \phi^2
  }^2
  \label{eq:characcoeff}
\end{eqnarray}
Since the coefficients $a,b,c$ in \eqref{eq:characcoeff} are explicitly background and reparametrization invariant, the mass squared singular values of the neutral singlet-doublet sector which are the solutions of the characteristic equation \eqref{eq:characeq} are also invariants. For completeness they are given by\begin{eqnarray}
 m_i^2 \! \! \! &=&\! \! \!  -\frac{1}{3 C}\bigg[~\! aC+\omega_iC^2+\frac{A}{\omega_i} ~\! \bigg]  \nonumber \\
  A\! \! \! &=&\! \! \!  a^2-3b  \nonumber \\
B\! \! \! &=&\! \! \!  2a^3 - 9ab + 27c \nonumber \\
  C\! \! \! &=&\! \! \!  \bigg[~\!  \frac{B}{2}+ \frac{1}{2}\sqrt{B^2 - 4A^3} ~\! \bigg]^{1/3}  \nonumber \\
  \omega_1\! \! \! &=&\! \! \!  1 \quad , \quad \omega_2=-\frac{1}{2}+i\frac{\sqrt{3}}{2} \quad , \quad \omega_3=\omega_2^*
  \label{eq:exactmasseigenstates}
\end{eqnarray}
with $i=1,2,3$. 

In this work, we are interested in studying the finite temperature effective Higgs potential, which determines the nature of the electroweak phase transition. Up to one-loop, the effective potential is determined by the tree level potential \eqref{eq:higgspontential1}, plus a zero-temperature and a finite temperature 1-loop contribution. The zero-temperature one-loop contribution is given by 
\begin{eqnarray}
V_\textrm{1-loop}
&\equiv&
\frac{1}{64 \pi^2}
~
\sum_{a}
(-1)^\xi
g_a
\,
\Big[
{m_{a}^4}
\Big(
\,
\log
\Big(
\,
\frac{m_{a}^2}{\mu^2}
\Big)
-
\frac{3}{2}
\Big)
\,
+P_a(\phi^2)
\Big]
\label{eq:zerotemp1-loop}
\end{eqnarray}
where $\mu$ is the renormalization scale, all couplings must be interpreted as effective couplings at that scale and $a$ is an index that runs over all boson and fermion fields obtaining mass from the Higgs mechanism. $\xi=1$ for fermions, $\xi=0$ for bosons. $g_a$ corresponds to the degrees of freedom of the corresponding field, which is equal to $1$ for a real scalar, $2$ for a Weyl fermion, $3$ for a neutral massive gauge boson. We only consider the contributions to the effective potential coming from the three new neutral Majorana fermions with masses specified in \eqref{eq:exactmasseigenstates}, from the W boson ($m_W=g_2 \phi/2$), the $Z$ boson ($m_Z=m_W\cos\theta_W$) and from the top quark ($m_t=y_t \phi/\sqrt{2}$). We neglect the subleading contributions coming from all the rest of the particles in the Standard Model. The functions $P_{a}(\phi^2)$ in \eqref{eq:zerotemp1-loop} depend on renormalization conditions, which are chosen to be
\begin{equation}
\frac{\partial}{\partial \phi}
\,
V_\textrm{1-loop}
\Big|_{\phi=v}
=
0
\qq
\qq
\frac{\partial^2}{\partial \phi^2}
\,
V_\textrm{1-loop}
\Big|_{\phi=v}
=
0
\label{eq:renormconditions}
\end{equation}
which up to a field independent term set the functions $P_{a}(\phi^2)$ to  \cite{Carena:2004ha,Davoudiasl:2012tu}
\begin{equation}
P_{a}(\phi^2)
=
\alpha_a
\phi^2
+
\beta_a
\phi^4
\label{eq:renormalizationfunction}
\end{equation}
\begin{equation}
\alpha_a 
=
\frac{1}{64\pi^2}
\bigg[
\Big(
-3 \frac{\omega_a \omega_a'}{v}
+ \omega_a'
+ \omega_a \omega_a''
\Big)
\Big(
\log \frac{\omega_a}{\mu^2}
-\frac{3}{2}
\Big)
-
\frac{3}{2}
\frac{\omega_a \omega_a'}{v}
+
\frac{3}{2}
\omega_a'^2
+
\frac{1}{2}
\omega_a
\omega_a''
\bigg]
\end{equation}

\begin{equation}
\beta_a 
=
\frac{1}{128\pi^2 v^2}
\bigg[
\,
2
\,
\Big(
 \frac{\omega_a \omega_a'}{v}
- \omega_a'
- \omega_a \omega_a''
\Big)
\Big(
\log \frac{\omega_a}{\mu^2}
-\frac{3}{2}
\Big)
+
\frac{\omega_a \omega_a'}{v}
-3
\omega_a'^2
-
\omega_a
\omega_a''
\bigg]
\end{equation}
where we defined
\begin{equation}
\omega_a =
m^2_a \Big|_{\phi=v}
\quad , \quad 
\omega_a' =
\frac{dm^2_a}{d \phi}  
\Big|_{\phi=v}
\quad , \quad
\omega_a'' 
=
\frac{d^2 m^2_a}{d \phi^2} 
\Big|_{\phi=v}
\,\,
\end{equation}
The renormalization conditions \eqref{eq:renormconditions} ensure that there is no explicit renormalization scale dependence in \eqref{eq:zerotemp1-loop} (up to a field independent term) and that up to one-loop, the electroweak symmetry breaking condition and Higgs boson mass expression are given by the usual tree level expressions
\begin{eqnarray}
\,
\frac{\partial}{\partial v}
\,
\Big[ V_\textrm{tree} + V_{\textrm{1-loop}} \Big]\,
\Big|_{\phi=v}
&=&
\sqrt{2} m^2 v+ \frac{\lambda }{\sqrt{2}}v^3
=
0
\label{eq:EWSB2}
\end{eqnarray}
\begin{eqnarray}
m_h^2
=(125 \, \textrm{GeV})^2
=
\frac{\partial^2}{\partial v^2}
\,
\Big[ V_\textrm{tree} + V_{\textrm{1-loop}} \Big]
\Big|_{\phi=v}
&=&
 m^2 +\frac{3}{2} \lambda v^2 = \lambda v^2
 \label{eq:higgsmass2}
\end{eqnarray}
where in the last equality of \eqref{eq:higgsmass2} we made use of \eqref{eq:EWSB2}. These relations set the tree level Higgs quartic defined in \eqref{eq:higgspontential1} to $\lambda=0.26$ and the Lagrangian mass to $m^2=-\frac{1}{2}\lambda v^2$. 

Finally, the one-loop, finite temperature correction to the Higgs effective potential is given by 
\begin{eqnarray}
\sum_{a=i,t,W,Z}
(-1)^\xi
\,
\frac{g_a T^4}{2\pi^2}
\int_0^{\infty}
dx \, x^2
\log \Big(
1 
-(-1)^\xi
\exp\Big[
-
\sqrt{x^2+ m_a^2/T^2}
\,\,\Big]
\Big)
\label{eq:thermal}
\end{eqnarray}
where again, $\xi=1$ for fermions, $\xi=0$ for bosons. $g_a$ corresponds to the degrees of freedom of the corresponding field, and for simplicity we only consider the contributions from the singlet-doublet neutral fermions, gauge bosons and the top quark. In the next section we will find that in the parameter space for which a strong first order phase transition is obtained, the critical temperature $T_c$ is always smaller than the mass of singlet-doublet fermions running in the loops, so we refrain from performing any high temperature expansion of the potential throughout this work. We leave for future investigations the effect of adding one-loop thermal masses to the bosons and fermions contributing to \eqref{eq:thermal}. Since at the critical temperature these corrections are of order $ \frac{1}{16 } \abs{\lambda_{u,d}}^2 T_c^2 < \abs{m_i^2}$ \cite{Chung:2009cb}, we do not expect them to modify our conclusions. The full temperature dependent effective potential is obtained by summing \eqref{eq:higgspontential1}, \eqref{eq:zerotemp1-loop} and \eqref{eq:thermal}.


\section{Strong first order phase transition from electroweak scale fermions}
\label{sec:phasetransition}

In this section we study the strength of the electroweak phase transition in the singlet-doublet model. The electroweak breaking condensate that minimizes the potential at the critical temperature $T_c$ is $\phi(T_c) \equiv v_c$. In what follows, we numerically determine the critical temperature $T_c$ of the electroweak phase transition and the strength of the phase transition $v_c/T_c$  from the full thermal effective potential given by the sum of  \eqref{eq:higgspontential1}, \eqref{eq:zerotemp1-loop} and \eqref{eq:thermal}. For simplicity and with the purpose of concentrating on the strength of the phase transition, in this section we limit ourselves to the CP conserving Singlet-Doublet model and postpone studying the effects of CP violation to sections  \ref{sec:CPviolating} and \ref{sec:results}. 
A sufficient condition for CP conservation in the Singlet-Doublet sector is $\delta_{\textrm{CP}}=0,\pi$, in which case and without loss of generality we may choose a field basis in which the Yukawas $\lambda_u, \lambda_d$ and the masses $m_S,m_L$ are real.

The scenario $\delta_{\textrm{CP}}=\pi$ corresponds to choosing three out of the four real Lagrangian parameters ($\lambda_u,\lambda_d,m_S,m_L$) to be positive and one negative. For this choice we find that level splitting only happens when $\abs{\lambda_u} \neq \abs{\lambda_d}$ and is insufficient. In particular, in the case $\abs{\lambda_u}=\abs{\lambda_d}$ the mass of the lightest neutral singlet-doublet fermion is either independent of or monotonically increasing with the Higgs field and the mechanism explained in the introduction is not realized. A numerical analysis confirms that no strong first order phase transition is found for the choice $\delta_{\textrm{CP}}=\pi$, so we do not study this case any further in this work.

For the rest of this section we concentrate in the case $\delta_{\textrm{CP}}=0$, where without loss of generality the Yukawas and Lagrangian masses may all be taken to be non-negative. We find that this case is a realization of the mechanism explained in the introduction leading to a barrier in the Higgs effective potential and to a strong first order phase transition.

\begin{figure}[htbp]
\begin{center}
\includegraphics[width=16.5cm]{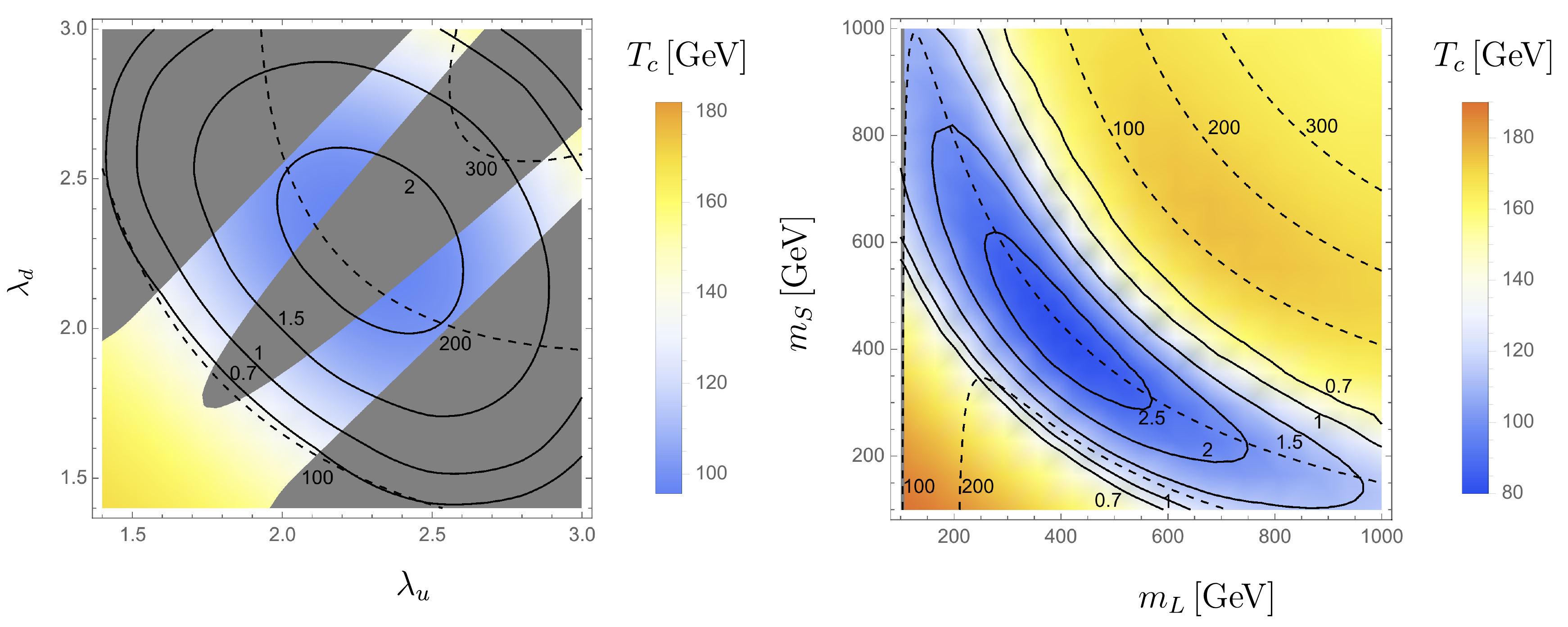}
\caption{\textit{Solid:} 
contours of the strength of the phase transition $\phi_c/T_c$, as a function of the singlet-doublet Yukawas $\lambda_{u,d}$ for $m_L=330 \, \textrm{GeV} \, ,  m_S=360 \, \textrm{GeV}$ (left), and as a function of ${m_{L,S}}$ for ${\lambda_u}=1.9 \, , \, {\lambda_d}=2.4$ (right). \textit{Dashed:} contours of the mass $m_1$ of the lightest singlet-doublet fermion in GeV.  \textit{Colored background:} density plot of the critical temperature $T_c$ of the electroweak phase transition.  \textit{Gray:} excluded by the electroweak precision analysis described in appendix \ref{app:EWP}.}
\label{fig:phasetransition}
\end{center}
\end{figure}

The results are shown in figure \ref{fig:phasetransition}, where in solid lines we plot contours of the strength of the phase transition $v_c/T_c$. We also show dashed contours of $m_1$, the mass of the lightest neutral fermion of the singlet-doublet sector and in the background we provide a colored density plot of the critical temperature $T_c$. In gray we show the areas excluded by electroweak precision constraints at $95\%$ confidence level according to the procedure described in appendix \ref{app:EWP}, which makes use of the STUVWX parameter formalism \cite{Maksymyk:1993zm}.
On the left panel of the figure, we first study the results as a function of the singlet-doublet Yukawas, where we fixed both Lagrangian masses to be close to the electroweak scale, $m_L= 330 \,  \textrm{GeV} \, , \,  m_S= 360 \,  \textrm{GeV}$. In this case, we see that a strong first order phase transition is obtained for Yukawa couplings in the range $1.5 \lesssim  \lambda_{u,d} \lesssim 3$. We also find that the critical temperature in the regions of parameter space where a strong first order phase transition occurs, is always smaller than the lightest singlet-doublet fermion mass $m_1$ as advertised in the introduction.  The cases $\lambda_{u,d} \gg \lambda_{d,u}$ generically do not lead to a strong first order phase transition: we find that the strength of the phase transition is maximized along the $\lambda_u = \lambda_d$ direction. Along this direction (or more generally, along the $\abs{\lambda_u} = \abs{\lambda_d}$ direction if we also allow for a physical CP violating phase), the singlet-doublet sector has an enhanced $SU(2)_R$ custodial symmetry which ensures a vanishing T parameter \cite{DEramo:2007anh,Enberg:2007rp,Cohen:2011ec,Abe:2014gua}. In spite of this, from the figure we see that for the order one Yukawas and non-decoupled singlet-doublet fermions needed for the strong first order phase transition, the $\lambda_u=\lambda_d$ direction is generically excluded, mostly due to a large S parameter. Moving slightly away from the $\lambda_u=\lambda_d$ direction leads to a small positive T parameter, which improves the electroweak precision fit for non-vanishing S (see figure \ref{fig:STfit}), avoiding thus the electroweak precision constraints. On the other hand, the regions with $\lambda_{u,d} \gg \lambda_{d,u}$ are excluded mostly due to a large T parameter. We conclude that generically, in order to avoid electroweak precision constraints while obtaining a strong first order phase transition, one needs to choose large (but perturbative) Yukawas $\lambda_u,\lambda_d$ and the two Yukawas must be similar. It is worth noting that the choice of similar singlet-doublet Yukawas arises quite naturally in a singlet-doublet sector which preserves custodial symmetry at the scale of some UV completion, in which case the singlet-doublet Yukawas at the electroweak scale would only be split by radiative custodial-breaking corrections. 

%
In figure \ref{fig:phasetransition} on the right, we study the results as a function of the singlet-doublet Lagrangian masses, where we fixed the Yukawas to $\lambda_u=1.9 \, , \,  \lambda_d=2.4$. We find that a strong first order phase transition is obtained for a large range of Lagrangian masses at the electroweak scale. For Lagrangian masses above $\sim 1 \, \textrm{TeV}$, the effects of the new fermions in the Higgs effective potential are Boltzmann suppressed at the scale of the electroweak phase transition and no strong first order transition is found. Of course, one could take even larger Yukawas, in which case the singlet-doublet masses could be as high as a few TeV as in \cite{Davoudiasl:2012tu}. However, as we will see in the next section, the baryon asymmetry is generated by reflection of the same set of singlet-doublet fermions on the bubble wall. Taking the singlet-doublet fermions much above the electroweak scale would suppress their abundance in the plasma at the critical temperature and would lead to a highly suppressed baryon asymmetry. On the opposite case, when both Lagrangian masses are smaller than $ \sim 300 \, \textrm{GeV}$ a strong first order phase transition is not achieved either. This can be understood by taking the limit $m_S, m_L \rightarrow 0$ in which case the neutral singlet-doublet fermions get mass only from the Higgs mechanism, no level splitting occurs, our mechanism is not realized and no barrier is created. Finally, note that the strength of the phase transition is maximal near $m_L=m_S$ and is left approximately unchanged upon exchange of $m_L$ and $m_S$.
This is a feature inherited from the ${\lambda_u} = {\lambda_d}$ custodial $SU(2)_R$ symmetric case, in which only two out of the three neutral fermions couple to the Higgs and the eigenvalues of the mass squared matrix entering the Higgs effective potential are exactly symmetric under $m_S \leftrightarrow m_L$. The small asymmetry under the exchange $m_L \leftrightarrow m_S$ in the right panel of figure \eqref{fig:phasetransition} is due to the small deviation from the $\lambda_u=\lambda_d$ case.


%

\subsection{Stability of the Higgs potential and Landau poles}
We found that in the singlet-doublet model, a strong first order phase transition requires large values of the Yukawas, $\lambda_u, \lambda_d $, as in the models presented in \cite{Carena:2004ha,Davoudiasl:2012tu}. This leads to an instability of the zero temperature Higgs potential below the TeV scale. In order to solve this problem, we introduce stabilizing irrelevant operators coming from a multi-TeV UV completion\begin{equation}
\frac{1}{\Lambda_n^{2n-4}} (H^\dagger H)^n
\label{eq:genericstabilizing}
\end{equation}
with $n\geq 3$. For illustration, in this section we consider a typical benchmark point with $\lambda_u=1.9, \lambda_d=2.4, m_S=330 \, \textrm{GeV}, m_L=360\textrm{GeV}$ which according to figure \ref{fig:phasetransition} leads to a strong first order phase transition. In this case, the Higgs potential instability is around $ \phi \approx 500 \textrm{GeV}$.  In order to stabilize the potential up to the cutoff of the theory, it suffices to add the operator
\begin{equation}
(H^\dagger H)^3/\Lambda^2
\label{eq:h6term}
\end{equation}
with cutoff $\Lambda \leq 1.2 \, \textrm{TeV}$. This stabilizing operator may be easily obtained from integrating out a multi-TeV scalar \cite{Egana-Ugrinovic:2015vgy}. The new scalar leads to a new tuned scale in the theory but in the multi-TeV range, where a UV completion which solves both the Higgs and new scalar hierarchy problems may be manifest. 

One may worry that the stabilizing operators \eqref{eq:genericstabilizing} affect the nature of the electroweak phase transition, either through the thermal effects of the underlying dynamics, or through its effect on the zero-temperature potential. However, if the underlying dynamics corresponds to a multi-TeV UV completion \textit{and} the cutoffs $\Lambda_n$ are much larger than the scale of the electroweak phase transition which is of the order of the electroweak scale $\sim v$, the thermal effects of the underlying dynamics are Boltzmann suppressed and are negligible, while the zero-temperature effects are suppressed by powers of $(v/\Lambda_n)^{2n-4}$. As a concrete example, for the benchmark point mentioned above with the stabilizing operator \eqref{eq:h6term} and $\Lambda=1.2 \, \textrm{TeV}$, we find that the  correction to the strength of the phase transition due to the stabilizing operator is less than $3\%$. This observation is quite general: we find that for all the Yukawas leading to a strong first order phase transition, one can always choose a multi-TeV UV completion leading to operators of the form \eqref{eq:h6term} such that the effects of the UV completion on the strength of the phase transition are at the percent level at most. This is a rather novel feature of our model, which ensures that the origin of the strong first order phase transition \textit{is entirely due to the new fermions in the theory} and extra multi-TeV scalars which may be the origin of the stabilizing operators  \textit{do not play any significant role} in either the formation of the barrier leading to the strong first order phase transition, nor on the calculation of the baryon asymmetry to be presented in the next section. 

%
%
%
%
%

Finally, the large Yukawas lead to Landau poles above the TeV scale. For the benchmark point above, using the one-loop beta functions given in appendix \ref{app:betafunctions}, we find that a Landau pole for the Yukawas is obtained at $\sim 40 \, \textrm{TeV}$. This also points to the need of building a UV completion for the theory, which is beyond the scope of this work.  Alternatives for the UV completion were already listed in \cite{Basirnia:2016szw}. They involve either providing a composite description of the model, or making copies of the singlet-doublet fields and promoting the corresponding multiplet fields to multiplets of a non-abelian gauge symmetry.

\FloatBarrier

\FloatBarrier
\section{CP violation in the singlet-doublet model}
\label{sec:CPviolating}
In this section we discuss the effects of CP violation in the singlet-doublet model. For our purposes, the two main features of considering a non-zero CP violating singlet-doublet phase are the generation of a baryon asymmetry during the electroweak phase transition and the generation of an electron electric dipole moment (EDM). Since there is a single effective CP violating phase in the singlet-doublet sector, both observables are related. We start by estimating the baryon asymmetry in \ref{sec:BAS} and in section \ref{sec:EDM} we present the limits from the electron EDM on the singlet-doublet effective phase.

\subsection{The baryon asymmetry: an analytic estimate}
\label{sec:BAS}
In this section we perform an approximate semiclassical calculation of the baryon asymmetry in the CP violating singlet-doublet model. We closely follow the techniques presented in \cite{Nelson:1991ab,Huet:1994jb,Huet:1995sh}.  More modern and sophisticated techniques exist to obtain the baryon asymmetry \cite{Riotto:1998bt,Riotto:1999yt}, but here we limit ourselves to a simpler but analytic estimate of the baryon asymmetry, in order to capture in a straightforward and intuitive way much of the physics that one would expect from a more precise calculation. 

We start by discussing the relevant timescales for the problem at temperatures close to the electroweak scale. The largest interaction rates correspond to the singlet-doublet Yukawa mediated processes, which for $\lambda_{u,d} \sim 2-3$, we estimate to be $10^{-2} \, T$ and the strong sphaleron rate which is of similar order \cite{Cline:1995dg}. The top Yukawa interaction is estimated to have a rate of $10^{-3} \, T$. For a wall velocity of $v_w=0.1$, quarks diffuse in front of the bubble wall at a rate of $10^{-3} \, T$, while leptons diffuse at a rate of $10^{-4} \, T$ \cite{Cline:1995dg}. Finally, electroweak sphalerons have a rate of $10^{-5} \, T$ \cite{Cline:1995dg}. We take all the rest of the Yukawa interactions in the Standard Model to be out of equilibrium and we neglect them in the rest of the calculation. 

This hierarchy of scales motivates the following simple picture for the production of the baryon asymmetry. First, an asymmetry in some global quantum number carried out by the vector-like doublets $\psi_L, \psi_{\bar{L}}$ is produced due to asymmetric scattering of the neutral components of the doublets on the bubble wall. Then, the fastest interaction rates, namely the strong sphalerons, singlet-doublet Yukawas and top quark Yukawas transform this vector-like doublet asymmetry into a chiral asymmetry for the Standard Model leptons and quarks. It turns out that this process is inefficient in the minimal singlet-doublet model, since strong sphalerons wash out most of the resulting chiral asymmetry in the model (like in the minimal supersymmetric standard model case, see \cite{Carena:2004ha,Huet:1995sh,Giudice:1993bb}), up to corrections inversely proportional to the strong sphaleron rate. This introduces an additional complication in the calculation of the baryon asymmetry. For the sake of brevity, we leave a detailed investigation of this issue for future work and in this paper we assume that all of the vector-like doublet asymmetry is efficiently transformed into a chiral asymmetry. This would be the case for instance if we allow the vector-like doublets to decay to standard model leptons and a new scalar or pseudoscalar (which must be odd under the $\mathbb{Z}_2 $ symmetry of table \ref{tab:gaugecharges}). The obtained chiral asymmetry then diffuses in front of the bubble wall for a distance equal to the mean free path of the fermions transporting the chiral asymmetry. Then, the slowest relevant process in the problem, namely the electroweak sphaleron interactions (which are active in front of the bubble wall), convert this chiral asymmetry into a baryon asymmetry, that eventually diffuses into the true vacuum inside the electroweak bubble. Finally, since the phase transition is strongly first order, the washout of the asymmetry inside the bubble by electroweak sphalerons is strongly suppressed and the comoving baryon asymmetry density remains unaltered for the rest of the evolution of the universe.

We start by providing an analytic calculation of the asymmetry created by reflection of the singlet-doublet fermions on the wall. First, we must define the global quantum number being created by asymmetric reflection on the wall. We choose this global quantum number to be ``doublet number" $U(1)_D$, under which $\psi_L$ has charge $+1$  $\psi_{\bar{L}}$ has charge $-1$. To ensure this number is approximately conserved and not washed out in the false vacuum, we also assign  $U(1)_D$ charge $-1$ to the Higgs and $+1$ to the doublet fermions of the SM, so doublet number is only violated by the slow down type Yukawa interactions, which we neglect \footnote{A more careful analysis requires finding the quantum number that is orthogonal to hypercharge in order to avoid Debye screening. We omit this technical detail which at most leads to a ${\cal O} (1)$ correction to the calculation \cite{Cohen:1992yh,Cline:1995di}.}. 

 In the thin wall approximation, where the bubble thickness $l$ is much smaller than the mean free path of the incoming fermions, the interactions with the bubble wall are captured by reflection and transmission coefficients of the incoming fermion wave. Since the bubble wall is macroscopic, we treat the reflection problem as one-dimensional and the singlet-doublet fermions as plane waves. In  \cite{Huet:1994jb,Huet:1995sh}, the reflection coefficients are calculated perturbatively from the Dirac equation and a simple interpretation for the result is provided, which we briefly summarize here. First, the fermions emerge from the thermal ensemble at some position which we define to be $z=0$. They propagate and reflect on the bubble wall a finite number of times, where each reflection in the perturbative calculation corresponds to one insertion of a space dependent fermion mass matrix. The bubble wall has a shape which we define to be 
 \begin{equation}
\phi(z)=\frac{1}{2} v_c \, \xi\Big(\frac{z-z_w}{l} \Big)
\label{eq:profiletemp}
\end{equation}
where $v_c$ is the critical condensate at the electroweak phase transition, $l$ the bubble width, $z_w$ the bubble wall position and $\xi$ is a dimensionless function which specifies the shape of the bubble and satisfies $\textrm{lim}_{x \rightarrow \infty} \, \xi(x) = 2 \, , \, \textrm{lim}_{x \rightarrow -\infty} \, \xi(x) =0$. 
 The space dependent mass matrix is obtained by using the vacuum profile \eqref{eq:profiletemp} in the mass matrix \eqref{eq:Mf} and is given by
\begin{equation}
{\mathcal M}(z)
=
\left(\begin{array}{ccc} m_S 
& 
\frac{1}{2\sqrt{2}} \lambda_u v_c \,\xi\big(\frac{z-z_w}{l} \big)
& 
\frac{1}{2\sqrt{2}} \lambda_d v_c \,\xi\big(\frac{z-z_w}{l} \big)
\\
 \frac{1}{2\sqrt{2}} \lambda_u v_c \,\xi\big(\frac{z-z_w}{l} \big)
 & 0 & 
 m_L \\ 
   \frac{1}{2\sqrt{2}} \lambda_d v_c \,\xi\big(\frac{z-z_w}{l} \big)& m_L & 0\end{array}\right)
\label{eq:mfprofile}
\end{equation}
The result of the perturbative calculation is an expression for the reflection and transmission coefficients as an expansion in powers of the fermion mass matrix ${\mathcal{M}}$ over the energy of the incoming fermions $\omega$, $
{\mathcal{M}}/{\omega}
$. The expansion for the  $3\times 3$ reflection coefficient matrix for incoming (right moving) singlet or doublet fermions from the unbroken phase into outgoing (left moving) singlet or doublet fermions is up to order ${\cal O} 
\big(
{\mathcal{M}^5}/{\omega^5}
\big)$ given by \cite{Huet:1994jb,Huet:1995sh}
\begin{eqnarray}
\nonumber
R
&=& 
\int_{0}^\infty
d {z_1}
\,
e^{2 (i\omega-\gamma) z_1}
{\cal M}^\dagger(z_1)
\,
\\
\nonumber
&&
~
+
~
\int_{0}^\infty
d {z_1}
\int_{z_1}^{0}
d {z_2}
\int_{z_2}^{\infty}
d {z_3}
\,
e^{2 (i \omega -\gamma)(z_1-z_2+z_3)} 
\,
{\cal M}^\dagger(z_1)
{\cal M}(z_2)
{\cal M}^\dagger(z_3)
\\
\nonumber
&&
~
+
~
\int_{0}^\infty
d {z_1}
\int_{z_1}^{0}
dz_2
\int_{z_2}^{\infty}
dz_3
\int_{z_3}^{0}
dz_4
\int_{z_4}^{\infty}
dz_5
e^{2(i \omega-\gamma) (z_1-z_2+z_3-z_4+z_5)}
\\
\nonumber
&&
\qq
{\cal M}^\dagger(z_1)
{\cal M}(z_2)
{\cal M}^\dagger(z_3)
{\cal M}(z_4)
{\cal M}^\dagger(z_5)
\\
&&
~
+
~
{\cal O} 
\Big(
\frac{\mathcal{M}^7}{\omega^7}
\Big)
\label{eq:RLR}
\end{eqnarray}
The parameter $\gamma$ in \eqref{eq:RLR} is a small damping term, which accounts for loss of coherence in the reflection due to interactions with the plasma and regulates the oscillatory integrals \eqref{eq:RLR}. We may understand the effect of $\gamma$ in the calculation by comparing it with the other two energy scales in \eqref{eq:RLR}: the fermion energy $\omega$ and the inverse bubble wall width $1/l$. First, we expect the damping rate to be of the order of the interaction rate with Higgs bosons in the plasma, due to the large singlet-doublet Yukawas needed to achieve the strong first order phase transition. For $\lambda_{u,d} \sim 2-3$, we estimate that these interactions have a rate $\sim 10^{-2}\,T$, which for a critical temperature of order $T_c \sim  100 \, \textrm{GeV}$ leads to $\gamma \sim 1 \, \textrm{GeV}$.  On the other hand, the energy of the incoming singlet-doublet fermions is of course larger than the singlet-doublet fermion masses, $\omega > \, \abs{m_{S,L}}$, which as discussed in section \ref{sec:phasetransition} are order electroweak scale. This means that $\omega \geq {\cal O} (10^2 \, \textrm{GeV}) \gg \gamma$. Finally, the width of the bubble wall is much harder to estimate reliably, since it involves the complex non-equilibrium evolution of the bubble on the plasma. A naive estimate of the bubble width may be obtained by minimizing the energy of the wall as in \cite{Cline:2006ts}, where the wall width is estimated to be $1/ l \sim 10^{2} \, \textrm{GeV}$. This calculation, however, does not account for the interactions of the wall with the plasma, so we will remain agnostic on the precise value of $l$ and in the rest of this work we treat the bubble wall width as a free parameter of order $10 \, \textrm{GeV} \lesssim 1/ l  \lesssim 10^{3} \, \textrm{GeV} $, in which case, $\gamma \ll 1/l$. To summarize, the damping rate is the smallest scale in \eqref{eq:RLR}, $\gamma \ll 1/l, \omega$. Then, to zeroth order in $\gamma l$ and $\gamma/\omega$, we may treat $\gamma$ just as a regulator of the oscillatory integrals  \eqref{eq:RLR}, which after integration may be set to zero. The error due to this approximation is of order ${\cal O}(\gamma l ,\gamma/\omega) \ll 1$. In the rest of this paper we omit writing $\gamma$ explicitly, with the implicit assumption that all oscillatory integrals are regulated as described.

The reflection matrix for the CP conjugate processes $\bar{R}$ is obtained by replacing the symmetric mass matrix ${\cal M}(z)$ in \eqref{eq:RLR} by its complex conjugate. The leading order term for the reflection asymmetry in doublet number arises at ${\cal O} \big( \mathcal{M}^6/\omega^6\big)$ and is given by
\begin{eqnarray}
\nonumber
\textrm{Tr}
\Big[
R^\dagger
\hat{Q}_D
R
-
\bar{R}^\dagger
\hat{Q}_D
\bar{R}
\Big]
&=&
4
\int_{0}^\infty 
d z_1
\int_{0}^\infty 
d z_2
\int_{z_2}^{0} 
dz_3
\int_{z_3}^{\infty} 
dz_4
\int_{z_4}^{0} 
dz_5
\int_{z_5}^{\infty} 
dz_6
\\
\nonumber
&&
\sin 2 \omega (z_1-z_2+z_3-z_4+z_5-z_6)
\\
\nonumber
&&
 \textrm{Im}
\,
\textrm{Tr}
\Big[
{\cal M}(z_1)
\hat{Q}_D
{\cal M}^\dagger(z_6)
{\cal M}(z_5)
{\cal M}^\dagger(z_4)
{\cal M}(z_3)
{\cal M}^\dagger(z_2)
\Big]
\nonumber 
\\
&+&
~
{\cal O} 
\bigg(
\frac{\mathcal{M}^8}{\omega^8}
\bigg)
\label{eq:RRdagger}
\end{eqnarray}
where the doublet number charge matrix is $\hat{Q}_D=\textrm{diag} \, (0,1,-1)$. Using the fermionic mass matrix \eqref{eq:mfprofile} in the doublet number reflection asymmetry \eqref{eq:RRdagger}, taking the trace of the matrices and performing a change of integration variables $z_i= x_i l$, $i=1..6$, we obtain
\begin{eqnarray}
\nonumber
\textrm{Tr}
\Big[
R^\dagger
\hat{Q}_D
R
-
\bar{R}^\dagger
\hat{Q}_D
\bar{R}
\Big]
&=&
\frac{\abs{m_S m_L}v_c^4}{8 \omega^6}
\abs{\lambda_u \lambda_d }
\big(
\lambda_u^* \lambda_u
-
\lambda_d^* \lambda_d
\big)
\,
\Xi(l  \omega)
\,
\sin \delta_{\textrm{CP}}
\\
&+&
~
{\cal O} 
\Big(
\frac{\mathcal{M}^7}{\omega^7}
\Big)
\label{eq:asymmetryreflection}
\end{eqnarray}
where we defined the dimensionless function $\Xi(l  \omega)$
\begin{eqnarray}
\nonumber
\Xi(l  \omega)
&=&
(l\omega)^6
\,
\int_{0}^\infty 
d x_1
\int_{0}^\infty 
d x_2
\int_{x_2}^{0} 
dx_3
\int_{x_3}^{\infty} 
dx_4
\int_{x_4}^{0} 
dx_5
\int_{x_5}^{\infty} 
dx_6
\\
\nonumber
&&
\sin \big[ 2 l \omega (x_1-x_2+x_3-x_4+x_5-x_6) \big]
\\
\qq
\,
&&
\xi(x_1) 
\Big[
\xi(x_3)\xi(x_5)
-\xi(x_2)\xi(x_4)
\Big]
\xi(x_6)
\label{eq:xifunction}
\end{eqnarray}
The function $\Xi(l  \omega)$ contains all the information of the bubble wall shape. Note that the reflection asymmetry \eqref{eq:asymmetryreflection} is independent of the wall position $z_w$. This result is valid as long as the singlet-doublet fermion emerges from the thermal ensemble far from the bubble wall, $z_w \gg l$. In this case the bubble wall position only leads to a phase $e^{2i \omega z_w}$ in the reflection coefficient, which does not affect the reflection probability. 

\begin{figure}[htbp]
\begin{center}
\includegraphics[width=10cm]{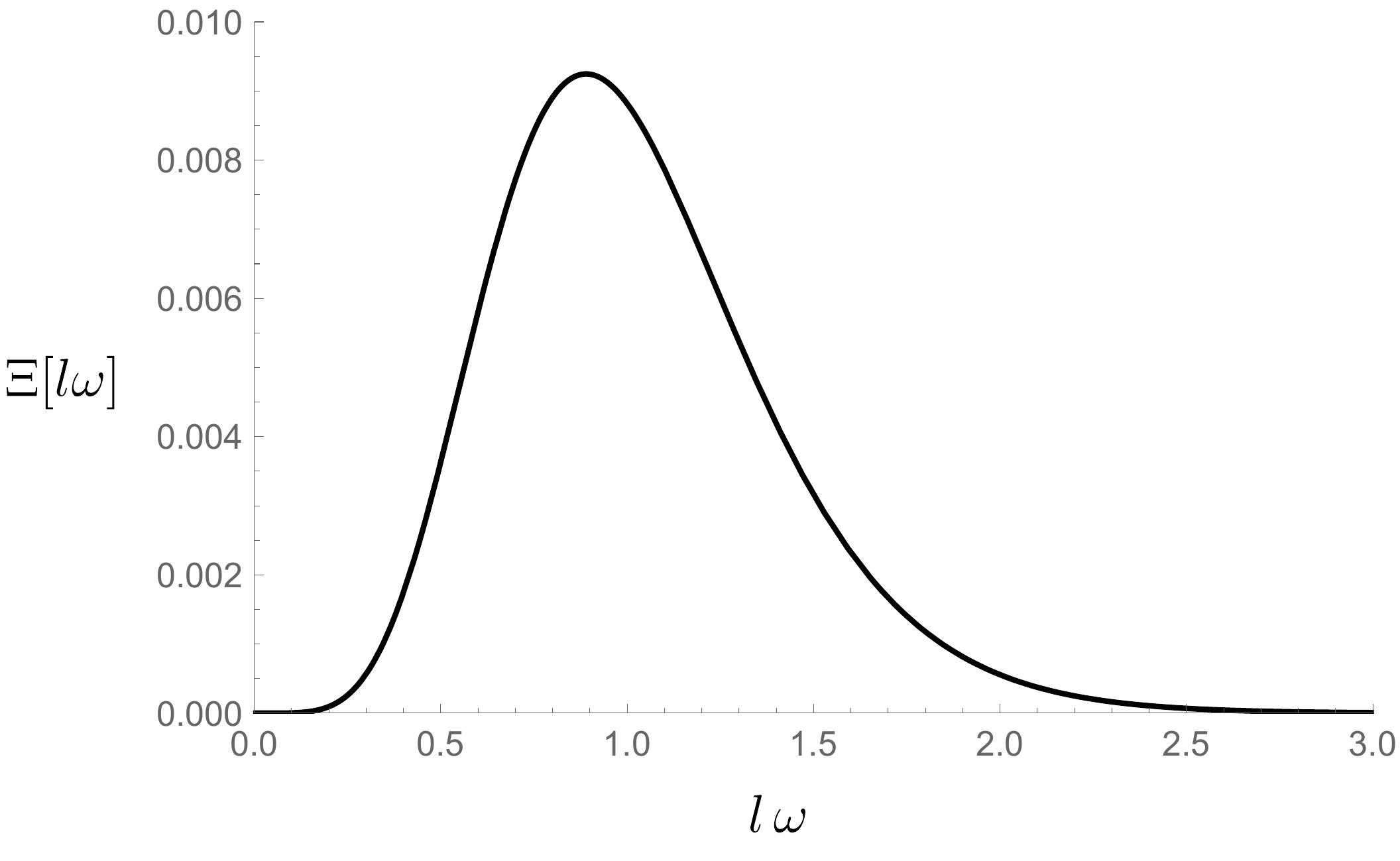}
\caption{Plot of the function $\Xi(l\omega)$ given in \eqref{eq:simplifiedxi}. $l$ is the bubble width and $\omega$ the energy of the incoming singlet-doublet fermion. The reflection asymmetry \eqref{eq:asymmetryreflection} is proportional to the function $\Xi(l\omega)$, which contains all the information of the bubble profile \eqref{eq:profile}, as may be seen from \eqref{eq:xifunction}.}
\label{fig:cascadeplot}
\end{center}
\end{figure}

In this work, for concreteness we take the bubble profile to have the usual kink shape \cite{Cline:2006ts,Cline:1995dg} which corresponds to the bubble shape function
\begin{equation}
\xi(x)= 1
+
\tanh x
\label{eq:profile}
\end{equation}
Inserting the bubble shape \eqref{eq:profile} in \eqref{eq:xifunction}, the function $\Xi(l  \omega)$ may be integrated analytically with some effort. The result is
\begin{eqnarray}
\nonumber
\Xi(l  \omega)
&=&
\frac{3}{32\pi^2}
(\pi l\omega)^3 
\,
\textrm{csch}^2
(\pi l\omega )
\,
\Big[
\,
\big(
\,
1
+
\pi l\omega \,
\textrm{coth}
(\pi l\omega )
\,
\big)
\,
\big(
\,
\gamma_E
+
~
\psi^0(-i l\omega)
\,
\big)
\\
& &
\qqqq \qqqq
+
i
l\omega
\psi^1(-i l\omega)
\Big]
+
\textrm{c.c.}
\label{eq:simplifiedxi}
\end{eqnarray}
%
%
where $\gamma_E=0.578$ is the Euler constant and $\psi^n(x)=d^{n+1}/dx^{n+1} \, \log \, \Gamma(x) $, $n=0,1$ are polygamma functions. For reference we plot $\Xi(l\omega)$ in figure \ref{fig:cascadeplot}. Inserting \eqref{eq:simplifiedxi} in \eqref{eq:asymmetryreflection} gives an analytic leading order expression for the reflection asymmetry of the vector-like doublet number. 

The transmission asymmetry for left moving singlet-doublet fermions coming from the broken phase may be similarly obtained from a perturbative calculation. However, it is simpler to obtain the transmission asymmetry by unitarity, which relates the reflection and transmission coefficients by
 \begin{equation}
\textrm{Tr}
\Big[
T^\dagger
\hat{Q}_D
T
-
\bar{T}^\dagger
\hat{Q}_D
\bar{T}
\Big]
=
-
\textrm{Tr}
\Big[
R^\dagger
\hat{Q}_D
R
-
\bar{R}^\dagger
\hat{Q}_D
\bar{R}
\Big]
\label{eq:Tasymmetry}
\end{equation}
The doublet number asymmetry in front of the wall may now be calculated in terms of the reflection and transmission asymmetries, but before proceeding and as a sanity check, let us consider some interesting limiting cases of the reflection asymmetry  \eqref{eq:asymmetryreflection}. First, the reflection asymmetry vanishes in the limit $\omega$ much larger than the wall height (which is controlled by $\lambda_{u,d}  \, v_c$ \,) as expected, since in this case the incoming singlet-doublet fermion has enough energy to penetrate in the bubble without reflecting. The asymmetry also vanishes when $m_S, m_L, \lambda_u$ or $\lambda_d$ are zero, since in this case there is no effective CP violation in the singlet-doublet model. Interestingly, in our leading order calculation, the reflection asymmetry vanishes when $\abs{\lambda_u} = \abs{\lambda_d}$, which corresponds to the custodial $SU(2)_R$ symmetric limit discussed in section \ref{sec:phasetransition}, but we do not expect this to hold at higher orders in the $\mathcal{M}/\omega$ expansion. Straightforward evaluation of the function $\Xi(l\omega)$ given in \eqref{eq:simplifiedxi} (or inspection of the plot in figure \ref{fig:cascadeplot}) indicates that there is a strong suppression of the reflection asymmetry both for $l\omega \gg 1$ and $l\omega \ll 1$. The limit $l\omega \gg 1$ corresponds to the case in which the quantum mechanical coherence needed for the reflection asymmetry is lost, due to interference from reflection at different points of the bubble profile \cite{Huet:1994jb}. The opposite limit $l\omega \ll 1$ corresponds to a zero thickness ``step wall", in which case all CP violation in the mass matrix in the broken vacuum may be rotated away by a unitary transformation \cite{Huet:1995mm}, so no asymmetry is created either. The reflection asymmetry is non-zero away from these two limits and is maximal for a bubble wall of thickness $l \sim 1/\omega$.

The doublet number asymmetry density in front of the bubble wall is given in terms of the doublet number reflection and transmission asymmetries by 
\cite{Huet:1994jb}
\begin{eqnarray}
\nonumber
n_D
&=&
T^2
\int_{\textrm{max}(m_S,m_L)}^\infty
\frac{d\omega}{2\pi}
\,
\Bigg[
~
\textrm{Tr}
\Big[
n^u(\omega)
\Big(
R^\dagger
\hat{Q}_D
R
-
\bar{R}^\dagger
\hat{Q}_D
\bar{R}
\Big)
\Big]
\\
&&
\qqqq
\qqqq
+
~
\textrm{Tr}
\Big[
n^b(\omega)
\Big(
T^\dagger
\hat{Q}_D
T
-
\bar{T}^\dagger
\hat{Q}_D
\bar{T}
\Big)
\Big]
~
\Bigg]
\label{eq:asymmetry0}
\end{eqnarray}
where $n^{u (b)}(\omega)$ is the unbroken (broken)  phase density matrix for the right (left) moving singlet and doublet fermions boosted to the wall frame. At lowest order in the expansion of mass over energy the density matrices are just proportional to the identity matrix \cite{Huet:1995sh}
\begin{eqnarray}
n^{u,b}(\omega)&=& \frac{1}{e^{\gamma_w (1\mp v_w )\omega/T_c }} \,  \textrm{diag} \, ( 1,1,1 ) 
+
{\cal O} 
\bigg(
\frac{\mathcal{M}}{\omega}
\bigg)
%
\label{eq:boltzmanndistr}
\end{eqnarray}
where $v_w$ is the bubble wall velocity, the minus sign is for the unbroken phase right moving fermions and the plus sign for the broken phase left moving fermions. $T_c$ is the temperature at which the baryon asymmetry is created, which we take to be the critical temperature for the electroweak phase transition. In this work we will not study the case of ultra-relativistic bubbles and we work at leading order in $v_w$. Using \eqref{eq:Tasymmetry} and \eqref{eq:boltzmanndistr} in \eqref{eq:asymmetry0} and expanding to first order in $v_w$ we obtain 
\begin{eqnarray}
\nonumber
n_D
&=&
2 v_w T_c^2
\int_{\textrm{max}(m_S,m_L)}^\infty
\frac{d\omega}{2\pi}
\,
n_0(\omega)
\big[
1-n_0(\omega)
\big]
\frac{\omega }{T_c}
\\
&&
\qqq
\frac{\abs{m_S m_L} v_c^4}{8 \omega^6}
\abs{\lambda_u \lambda_d }
\big(
\lambda_u^* \lambda_u
-
\lambda_d^* \lambda_d
\big)
\,
\Xi(l  \omega)
\,\sin \delta_{\textrm{CP}}
\\
\nonumber
&& 
\qqq
\Bigg[
1
+
{\cal O}\bigg( v_w^2, \frac{\mathcal{M}^7}{\omega^7} \bigg)
\Bigg]
\label{eq:asymmetry}
\end{eqnarray}
where $n_0(\omega)=(e^{\omega/T_c}+1)^{-1}$ is the Fermi-Dirac distribution.  

As already discussed in the beginning of this section, we assume that all the vector-like doublet asymmetry $n_D$ is efficiently converted into the chiral asymmetry density $n_L(x)$ in front of the wall, where $x$ is the distance from the wall. For simplicity, we assume that the chiral asymmetry density $n_L(x)$ is constant and equal to $n_D$ up to a distance $\Delta$ from the wall and zero beyond that distance, where $\Delta$ is the mean free path of the fermions transporting the chiral asymmetry,
\begin{gather}
n_L(x) =
\begin{cases}
 n_D    & \textrm{for } x \le \Delta \\
  0    & \textrm{for } x > \Delta
\end{cases}
\label{eq:nL}
\end{gather}
We take the mean free path to be $\Delta=100/T$, which is the mean free path of the SM leptons \cite{Cline:1995dg,Joyce:1994bi,Joyce:1994zn}, motivated by the possibility discussed in the beginning of this section that the singlet-doublet fermions may decay to Standard Model leptons, such that the chiral asymmetry is a lepton asymmetry. The baryon asymmetry is obtained from the space-dependent chiral asymmetry $n_L(x)$ which biases weak sphaleron interactions and is given by \cite{Cline:1995dg}
\begin{equation}
n_B=
-\frac{9}{T_c^3 v_w}  
\Gamma_{\textrm{sph}}
\int_{0}^{\infty}
d x
\,
n_L(x)
+
{\cal O}
\Bigg(
\frac{\Gamma_{\textrm{sph}}^2 \Delta^2 n_D}{v_w^2 T_c^6}
\Bigg)
\label{eq:dnbdt}
\end{equation}
where the weak sphaleron rate per unit volume at the electroweak phase transition is 
\begin{equation}
\Gamma_{\textrm{sph}}= \kappa (\alpha_W T_c)^4
\label{eq:sphaleronrate}
\end{equation}
and we take $\kappa=1.1$ \cite{Ambjorn:1995xm}. Using \eqref{eq:nL}  in \eqref{eq:dnbdt} we obtain 
\begin{eqnarray}
n_B
&=&
-\frac{9 \Delta}{T_c^3 v_w}\Gamma_{\textrm{sph}} 
n_D
\Bigg[
1
+
{\cal O}
\Bigg(
\frac{\Gamma_{\textrm{sph}}^2 \Delta^2}{v_w^2 T_c^6}
\Bigg)
\Bigg]
\label{eq:nBtemp}
\end{eqnarray}
Finally, using \eqref{eq:asymmetry} in \eqref{eq:nBtemp} we get
\begin{eqnarray}
\nonumber
n_B
&=&
-\frac{18 \Delta}{T_c}\Gamma_{\textrm{sph}} 
\int_{\textrm{max}(m_S,m_L)}^\infty
\frac{d\omega}{2\pi}
\,
n_0(\omega)
\big[
1-n_0(\omega)
\big]
\frac{\omega }{T_c}
\\
&&
\qqq
\frac{m_S m_L v_c^4}{8 \omega^6}
\abs{\lambda_u \lambda_d }
\big(
\lambda_u^* \lambda_u
-
\lambda_d^* \lambda_d
\big)
\,
\Xi(l  \omega)
\,\sin \delta_{\textrm{CP}}
\\
\nonumber
&&
\qqq
\Bigg[
1
+
{\cal O}
\Bigg(
v_w,\frac{\mathcal{M}^7}{\omega^7} , \frac{ \Gamma_{\textrm{sph}}^2 \Delta^2}{v_w^2 T_c^6}
\Bigg)
\Bigg]
\label{eq:nB}
\end{eqnarray}
where we remind the reader that the function $\Xi(l\omega)$ is given in expression \eqref{eq:simplifiedxi}, $n_0(\omega/T_c)$ is the Fermi-Dirac distribution, $\Gamma_\textrm{sph}$ is the sphaleron rate \eqref{eq:sphaleronrate} and $\Delta=100/T$. Expression \eqref{eq:nB} is a leading order, analytic, background symmetry invariant estimation of the baryon asymmetry and is the main result of this section.  The critical temperature and critical condensates at the electroweak phase transition $T_c, v_c$ are numerically determined from the finite temperature Higgs effective potential as described in section \ref{sec:phasetransition}. The baryon asymmetry \eqref{eq:nB} depends on all the five CP invariants of the singlet-doublet model described in section \ref{sec:model}, namely the CP even invariant magnitudes of the singlet-doublet Yukawas $\abs{\lambda_u},\abs{\lambda_d}$ and Lagrangian masses $\abs{m_S}, \abs{m_L}$, and the CP odd invariant phase $\delta_{\textrm{CP}}$ defined in equation \eqref{eq:definitiondelta}. The baryon asymmetry vanishes if any of these parameters is zero, since in this case there is no CP violation in the singlet-doublet model. 
The baryon asymmetry also depends on the bubble wall width $l$, but is independent within our approximation of the bubble velocity $v_w$. The approximation is valid as long as the weak sphaleron rate may be considered to be slow with respect to the expansion of the bubble, $\Gamma_{\textrm{sph}} \Delta/T^3 < v_w$. For very slow bubbles, $v_w \ll \Gamma_{\textrm{sph}} \Delta/T^3$, the baryon asymmetry washout due to electroweak sphalerons  in the unbroken phase must be included in the calculation and expression $\eqref{eq:nBtemp}$ needs to be replaced by $n_B=- 9 n_D \Big[ 1- \exp(-\Gamma_{\textrm{sph}}\Delta /(T^3 v_w))  \Big] $, which vanishes in the limit $v_w \rightarrow 0$ (since $n_D$ is linear in $v_w$, see eq. \eqref{eq:asymmetry}). This is to be expected, since for a static bubble the system is in equilibrium and no baryon asymmetry can be generated. For ultra-relativistic bubbles, our lowest order velocity expansion breaks down.  In what follows we stick to the case $ \Gamma_{\textrm{sph}} \Delta/T^3 < v_w \ll 1 $ and work with expression \eqref{eq:nB}

\subsection{The electron electric dipole moment}
\label{sec:EDM}
A singlet-doublet phase $\delta_{\textrm{CP}}\neq 0, \pi$ leads to an electron EDM through two-loop Barr-Zee diagrams \cite{Barr:1990vd}. The two loop diagrams were calculated in  \cite{Abe:2014gua}, results that we use to set limits on the effective singlet-doublet CP violating phase by comparing with the experimental limits on the electron EDM \cite{Baron:2013eja}.

\begin{figure}[htbp]
\begin{center}
\includegraphics[width=13.5cm]{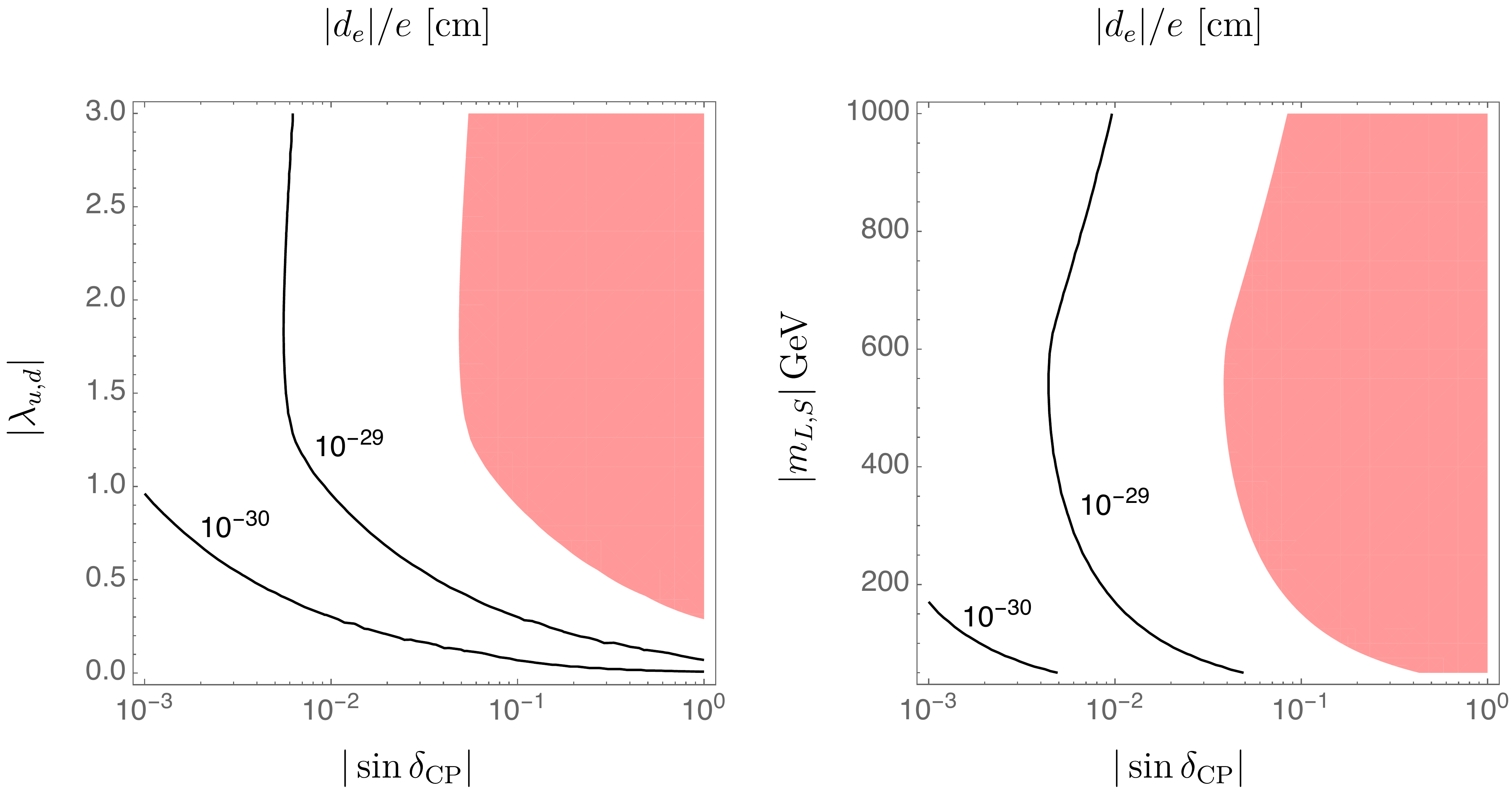}
\caption{\textit{Solid}: contour plots of the absolute value electron electric dipole moment in the singlet-doublet model as a function of $\abs{\sin \delta_{\textrm{CP}}}$ and equal singlet-doublet Yukawas $\abs{\lambda_u}=\abs{\lambda_d}$ for $\abs{m_L}=\abs{m_S}=300\,\textrm{GeV}$ (left) and as a function of $\abs{\sin \delta_{\textrm{CP}}}$ and equal singlet-doublet model Lagrangian masses $\abs{m_L}=\abs{m_L}$ for $\abs{\lambda_u}=\abs{\lambda_d}=2.5$ (right). \textit{Red}: region excluded by the electron EDM limit \cite{Baron:2013eja}, $\abs{d_e}/e \leq 8.7 \times 10^{-29} \, \textrm{cm}$.}
\label{fig:edmplot}
\end{center}
\end{figure}

In figure \ref{fig:edmplot} we present the limits on the (absolute value) sine of the CP violating phase \eqref{eq:definitiondelta}, $\abs{\sin \delta_{\textrm{CP}}}$. On the left, we present the exclusion region as a function of the absolute value of the singlet-doublet Lagrangian masses, by setting them equal for simplicity, $\abs{m_L}=\abs{m_S}$ and by fixing the absolute value of the Yukawas at $\abs{\lambda_u}=\abs{\lambda_d}=2.5$. We also include contours of the electric dipole moment $d_{e}/e$. We generically find that for such Yukawas, for $200 \, \textrm{GeV} \lesssim \abs{m_L}=\abs{m_S} \lesssim 900\, \textrm{GeV}$, the effective phase is constrained to be at the percent level or below. The limits degrade at lower masses, since in the limit of vanishing Lagrangian masses there is no effective CP violation in the singlet-doublet model. On the right plot we present the limits as a function of the absolute value of the Yukawas, which for simplicity are set to be equal $\abs{\lambda_u}=\abs{\lambda_d}$, for fixed singlet-doublet lagrangian masses, $\abs{m_L}=\abs{m_L}=300\,\textrm{GeV}$. For $\abs{\lambda_{u,d}} \gtrsim 1$, the singlet-doublet CP violating phase is again constrained to be at the percent level. The electron EDM is roughly independent of the absolute value Yukawas for $\abs{\lambda_{u,d}} \gtrsim 1$. Generically, we conclude that for electroweak-scale singlet-doublet masses and order one Yukawas close to the custodial preserving limit $\abs{\lambda_u} = \abs{\lambda_d}$, the singlet-doublet effective CP violating phase, $\delta_{\textrm{CP}}$ is constrained to be at most at the few percent level.

\section{Putting all together: the baryon asymmetry in the singlet-doublet model}
\label{sec:results}

The baryon asymmetry in the singlet-doublet model \eqref{eq:nB} needs to be compared with the measured value of the baryon asymmetry of the universe, which is given by \cite{Ade:2015xua} 
\begin{equation}
\frac{\abs{n_{B \, , \, \textrm{obs}}}}{s}=(8.6\,  \pm \, 0.09) \times 10^{-11}
\label{eq:expnb}
\end{equation}
where $s=g_* (2\pi^2/45)T^3$ is the entropy density and $g_*$ is the number of degrees of freedom in thermal equilibrium in the plasma, which we take to be the SM degrees of freedom plus the singlet-doublet fermions, $g_*=115.5$. The sign of the asymmetry is not determined in the measurement \cite{Ade:2015xua}, but for practical purposes we assume that it corresponds to a positive baryon asymmetry. 

\begin{figure}[htbp]
\begin{center}
\includegraphics[width=16cm]{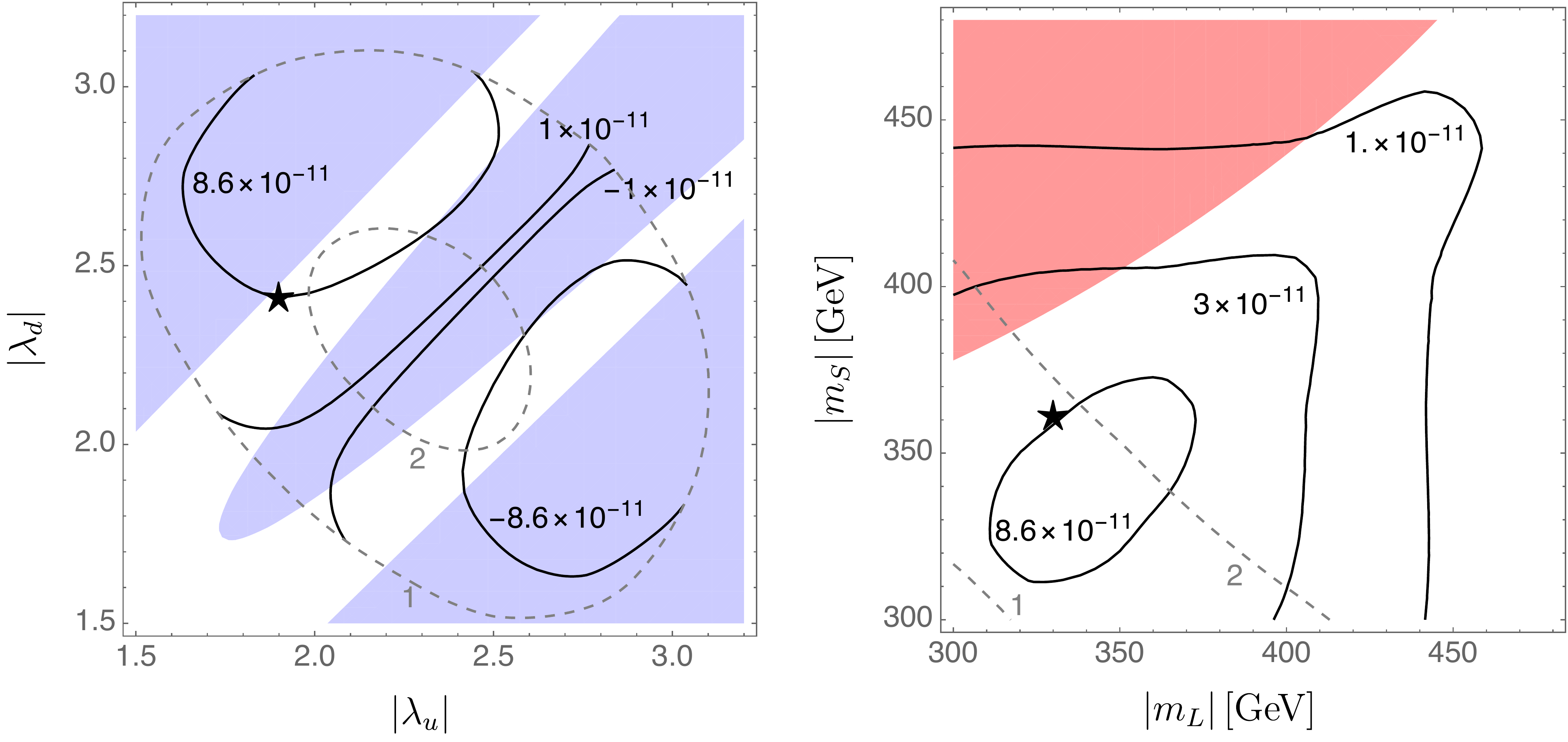}
\caption{
\textit{Solid:} 
contours of the baryon asymmetry over entropy density $n_B/s$ in the singlet-doublet model, as a function of the absolute value of the singlet-doublet Yukawas $\abs{\lambda_{u,d}}$ for $\abs{m_L}=330 \, \textrm{GeV} \, ,  \,\abs{m_S}=360 \, \textrm{GeV}$ (left) and as a function of $\abs{m_{L,S}}$ for $\abs{\lambda_u}=1.9 \, , \, \abs{\lambda_d}=2.4$ (right). In both panels the effective CP violating phase is $\delta_{\textrm{CP}}=- 4 \times 10^{-2}$ and the bubble wall width is set to $l=3\times 10^{-3} \, \textrm{GeV}^{-1}$. \textit{Dashed:} contours of the strength of the phase transition $v_c/T_c$. \textit{Blue:} excluded by the electroweak precision analysis described in appendix \ref{app:EWP}. \textit{Red:} excluded by the experimental limit on the electron EDM,  $d_e \geq 8.7 \times 10^{-29} \, \textrm{e cm}$ \cite{Baron:2013eja}. \textit{Star:} benchmark scenario of table \ref{tab:benchmark2}.
}
\label{fig:baryonasymmetry}
\end{center}
\end{figure}

The results are shown in figure \ref{fig:baryonasymmetry}, where we plot contours of the baryon asymmetry \eqref{eq:nB} over the entropy density, $n_B/s$. In the plots we fix the effective CP violating phase to $\abs{\delta_{\textrm{CP}}}= 4 \times 10^{-2}$, which is basically close to the maximum phase allowed by the electron EDM, as discussed in section \eqref{sec:EDM}, for electroweak scale singlet-doublet masses and Yukawas larger than one. The sign of the phase is not relevant for the EDM limit discussed in section \eqref{sec:EDM}, but is in principle measurable in a low energy experiment and is correlated with the sign of the baryon asymmetry, and for the plots we set the sign of the phase to be negative. We also set the bubble wall width to $l=3\times 10^{-3} \, \textrm{GeV}^{-1}$, which is of the order of the estimate in \cite{Cline:2006ts}. In the plots we also show contours of the strength of the phase transition $v_c/T_c$ in dashed gray lines. Note that we cut the contours of the baryon asymmetry  \eqref{eq:nB} in the regions of parameter space where the strength of the phase transition is less than one, since in that region the baryon asymmetry is washed out by weak sphalerons. Finally, we show in blue and red the regions excluded by the electroweak precision constraints (see appendix \ref{app:EWP}) and by the electron EDM limits discussed in section  \ref{sec:EDM}. 

On the left of figure \ref{fig:baryonasymmetry}, the results are shown as a function of the absolute value of the singlet-doublet Yukawas $\abs{\lambda_u}, \abs{\lambda_d}$ for fixed Lagrangian masses $\abs{m_S}=360 \, \textrm{GeV}, \abs{m_L}=330 \, \textrm{GeV}$. We see that the model is able to reproduce the baryon asymmetry for Yukawas of order $2 \lesssim \abs{\lambda_{u,d}} \lesssim 3$. The star shows a typical benchmark point, presented in table \ref{tab:benchmark2}, which leads to a strong first order phase transition, reproduces the baryon asymmetry and avoids electroweak precision and EDM constraints. We postpone commenting on the collider constraints for this benchmark scenario to section \ref{sec:collider}. Note that the baryon asymmetry vanishes along the $SU(2)_R$ custodial preserving limit $\abs{\lambda_u} = \abs{\lambda_d}$ and the sign of the asymmetry is opposite for $\abs{\lambda_d} > \abs{\lambda_u}$ and $\abs{\lambda_u} > \abs{\lambda_d}$. Both these features are an artifact of keeping only the lowest order term in the mass expansion in the calculation of the baryon asymmetry of section \ref{sec:BAS}. We do not expect these features to survive at higher order in the mass expansion and we leave the quantification of the departure from the lowest order calculation for future work. On a different note, from a naive analysis of the baryon asymmetry \eqref{eq:nB}, one would expect it to grow monotonically with larger Yukawas. This is not the case, since if the Yukawas are too large, the singlet-doublet fermions are heavy and their thermal effects on the Higgs effective potential are Boltzmann suppressed. In this case, both the strength of the phase transition $v_c/T_c$ and the Higgs critical condensate $v_c$ are suppressed. This not only leads to suppression by washout from electroweak sphalerons in the broken phase, but also decreases the height of the bubble wall, which is controlled by $v_c$, suppressing the singlet-doublet reflection asymmetry and therefore the baryon asymmetry (the dependence of the baryon asymmetry on the wall height may be explicitly seen in $\eqref{eq:nB}$, $n_B \propto v_c^4$). 

The plot on the right of figure \ref{fig:baryonasymmetry} shows the baryon asymmetry as a function of the Lagrangian singlet-doublet masses for fixed Yukawas, $\abs{\lambda_d}=2.4, \abs{\lambda_u}=1.9$. We see that generically the baryon asymmetry of the universe can be reproduced only for electroweak-scale singlet-doublet Lagrangian masses. The baryon asymmetry is suppressed both in the small and large mass regions, in the first case due to the suppression of the strong first order phase transition and suppression of the effective CP violating invariant, and in the latter case mostly due to Boltzmann suppression of the singlet-doublet abundance in the plasma. The star indicates again the benchmark point of table \ref{tab:benchmark2}.

Generically, we conclude that in order to obtain the strong first order phase transition and to obtain the observed baryon asymmetry from reflection of singlet-doublet fermions on the bubble wall, the singlet doublet fermions must have large Yukawas with the Higgs and be at the electroweak scale. We expect these features to be generic in all models of fermion induced electroweak baryogenesis, since they have a generic origin: the large Yukawas are needed in order to substantially modify the Higgs potential at the electroweak phase transition, while the new fermions must be below a TeV to avoid suppressing their abundance from the thermal plasma at the electroweak phase transition.

\begin{table}[ht!]
\begin{center}
$$
\begin{array}{|c|c||c|c|} 
\hline
\abs{m_S} 
&
360 \, \textrm{GeV}
&
m_1
&
 172.8, \textrm{GeV}
\\ \hline
\abs{m_L} 
&
330 \, \textrm{GeV}
&
m_2
&
343.2 \, \textrm{GeV}
\\ \hline
\abs{\lambda_u} 
&
1.9
&
m_3
&
875.7 \, \textrm{GeV}
\\ \hline
\abs{\lambda_d} 
&
2.4
&
 \delta_\textrm{CP}
&
- 4 \times 10^{-2}
\\
\hline
\end{array}
$$
\end{center}
\caption{Benchmark scenario shown with a star in figure \ref{fig:baryonasymmetry}. This example point leads to a strong first order phase transition, reproduces the baryon asymmetry of the universe in the leading order estimate \eqref{eq:nB} and avoids electroweak precision and EDM constraints. Note that all the singlet-doublet fermions are close to the electroweak scale.}
\label{tab:benchmark2}
\end{table}

\section{Collider constraints}
\label{sec:collider}
In this section we briefly comment on the collider constraints on the charged and neutral components of the minimal singlet-doublet model. 

We start reviewing the limits for singlet-doublet fermion masses below $100 \textrm{GeV}$. The irreducible limit on the mass $\abs{m_L}$ of the charged component of the doublet fermions is basically half the Z boson mass \cite{Heister:2002mn}. There are stronger constraints from direct pair production at LEP under certain assumptions for the decay modes and lifetime of the charged fermion, which set a stronger bound $\abs{m_L} \gtrsim 91\textrm{GeV}$ \cite{lepsusy}. There are limits on the masses of the neutral singlet-doublet fermions if they are below half the Z boson mass from the Z invisible width, but it is not irreducible. For instance, in the custodial symmetric limit the lightest neutral singlet-doublet fermion does not couple to the Z. There is a more important limit on neutral singlet-doublet fermions if they have a Yukawa coupling to the Higgs of the order of the bottom Yukawa or larger, coming from the Higgs invisible width measurement \cite{Khachatryan:2016whc,Aad:2015pla}, which basically rules out neutral singlet-doublet fermions below half the Higgs mass. The high luminosity LHC will be able to probe pair production of neutral singlet-doublet fermions if they are mostly doublet up to $\sim 100 \, \textrm{GeV}$ even in the compressed region, from searches assisted by the emission of an initial state radiation jet \cite{Schwaller:2013baa}.

In the larger mass region, with singlet-doublet fermions heavier than $100 \, \textrm{GeV}$, the main collider constraints come from LHC, where singlet-doublet fermions are pair produced and subsequently decay into the lightest fermion of the singlet-doublet sector, leading to final states with missing energy, multiple leptons and/or jets. The strongest constraints come from searches with multiple leptons in the final state, both at CMS \cite{CMS:2017fdz,CMS:2016gvu,CMS:2016zvj} and ATLAS \cite{ATLAS:2016uwq,ATLAS:2017uun}. This is the region of interest for our benchmark scenario of table \ref{tab:benchmark2}, which we now discuss in more detail.

To study our benchmark scenario, we implement the minimal singlet-doublet fermion model using \textsc{FeynRules} \cite{Alloul:2013bka}. The relevant topologies for LHC are pair production of the charged fermions $\psi_+ \psi_-$, or production of a charged fermion and a neutral fermion, $\psi_\pm \psi_i$, $i=1,2,3$. The charged fermions are produced through Drell-Yan and the neutral singlet-doublet fermions are produced either through Drell-Yan or through an s-channel Higgs \cite{Basirnia:2016szw}. Since in our benchmark scenario, the third neutral state $\psi_3$ is more than a factor of two heavier than the rest of the singlet-doublet fermions, its production is subleading and we concentrate on production modes involving only $\psi_\pm$ and $\psi_{1,2}$. For simplicity, we also limit ourselves to pair production $\psi_\pm \psi_2$ and $\psi_{+} \psi_-$, which are the most similar topologies to the ones in \cite{CMS:2017fdz,CMS:2016gvu,CMS:2016zvj,ATLAS:2016uwq,ATLAS:2017uun}, but including also production of $\psi_1\psi_2$ does not change the conclusions. We use \textsc{Madgraph} \cite{Alwall:2014hca} to perform a  Montecarlo simulation for pair production at the 13 TeV LHC and to obtain the decay branching fractions of the singlet-doublet fermions. We tabulate the resulting production cross sections and main decay modes in \ref{tab:BRs}. Regarding the decay modes, the charged fermion $\psi_\pm$ decays to a W boson and the lightest singlet-doublet neutral fermion, which is the lightest stable particle in the minimal singlet-doublet sector, leaving missing energy and leptons or jets. The second heaviest singlet-doublet neutral fermion $\psi_2$ decays to a Z or Higgs and the lightest neutral singlet-doublet fermion. We find that the decay through the Higgs is dominant. This is a typical feature of models of fermion induced electroweak baryogenesis, due to the requisite large Yukawa couplings to the Higgs and provides motivation for searches with Higgs mediated decays as in \cite{CMS:2017fdz}.  Note that the $\psi_2$ decay to a charged doublet fermion and a $W^*$ is three body, so it is suppressed. 

We subsequently recast the $13 \, \textrm{TeV}$ ATLAS limits \cite{ATLAS:2016uwq} using \textsc{CheckMate} \cite{Dercks:2016npn}.  The corresponding CMS multilepton searches \cite{CMS:2017fdz,CMS:2016gvu} are not currently implemented in \textsc{CheckMate}, but the limits are expected to be similar. We find that our benchmark scenario is not currently excluded, for rather trivial reasons. For the kinematics of our benchmark point, the limit on the production cross section times branching fraction into W and Z bosons for the pair $\psi_\pm \psi_2$  is ${\cal O} (0.1 \textrm{pb})$ according to \cite{CMS:2017fdz}. This limit is one to two orders of magnitude larger than the corresponding production cross section times branching fraction for our benchmark point (see table \ref{tab:BRs}). However, the limits will improve considerably with more statistics, so the benchmark scenario may be tested at the high luminosity LHC. We leave a detailed study of the full parameter space of the singlet-doublet model and the discovery prospects for future work. 

Another possibility, is to allow for singlet-doublet fermions to decay to the Standard Model lepton doublet and a new light scalar or pseudoscalar singlet, as discussed in section \ref{sec:BAS}. In this case, the charged singlet-doublet fermion decays mostly into a Standard Model charged lepton, while the singlet-doublet fermions decay invisibly. In this scenario, the strongest constraints come from pair production of charged singlet-doublet fermions and depend mostly on the mass splitting between the charged singlet-doublet fermion and the new scalar or pseudoscalar. We also leave a detailed study of this interesting possibility for future work.

\begin{table}[ht!]
\begin{center}
$$
\begin{array}{|c|c||c|c|} 
\hline
\sigma_{\psi_+ \psi_-}
&
1.3 \times 10^{-2}\, \textrm{pb}
&
\textrm{BR}(\psi_\pm \rightarrow W^\pm \psi_1)
&
1
\\ \hline
\sigma_{\psi_\pm \psi_2}
&
6.9 \times 10^{-3}\, \textrm{pb}
&
\textrm{BR}(\psi_2 \rightarrow Z \psi_1)
&
0.45
\\ \hline

&
&
\textrm{BR}(\psi_2 \rightarrow h \psi_1)
&
0.55
\\ \hline
\end{array}
$$
\end{center}
\caption{Pair production cross sections for the main modes at the $13 \, \textrm{TeV}$ LHC leading to multiple leptons in the final state and branching fractions of the charged and neutral singlet-doublet fermions for the benchmark scenario of table \ref{tab:benchmark2}. The Singlet-Doublet model is implemented with \textsc{FeynRules} \cite{Alloul:2013bka} and cross sections and branching fractions are obtained with \textsc{Madgraph} \cite{Alwall:2014hca}.}
\label{tab:BRs}
\end{table}

\section{Conclusions and outlook}
\label{sec:conclusions}

We presented the minimal model of electroweak baryogenesis induced only by fermions. We demonstrated that the model is complete: it leads to a barrier in the effective potential induced entirely by the new fermions and leads to a baryon asymmetry of the correct order of magnitude in a simplified semiclassical calculation, which was performed analytically in terms of CP invariants. We showed that in order for the strong first order phase transition and baryon asymmetry to be generated, the singlet-doublet fermions must be at the electroweak scale, have large Yukawas with the Higgs and have non-vanishing Lagrangian masses in order to realize level splitting in the broken vacuum.  We also studied the most relevant experimental constraints, including electroweak precision, electron dipole moment and collider constraints, showing that the model is consistent with current experimental data. 

Fermion induced EWBG is largely unexplored in the literature in comparison with scalar induced EWBG and poses many interesting questions. First, a detailed study of the phenomenology of fermion induced EWBG is not available. Such a study should include a full analysis of the irreducible collider constraints and discovery prospects at LHC, which must include a full recast of multilepton searches. Fermion induced EWBG is a very well motivated model of new physics at the electroweak scale with strong couplings with the Higgs, so on its own it is an interesting benchmark scenario for LHC. Also, a study of the deviation of the self couplings must be carried out, but we point out that this study is sensitive to the UV completion stabilizing the potential.  

On the technical side, a more precise calculation of both the phase transition and the baryon asymmetry in fermion induced EWBG should be carried out. Due to the large Yukawas, it would be interesting to study the effects of including higher loop terms in the effective potential for the calculation of the strength of the phase transition. Also, in this work we neglected the effect of washout of the chiral asymmetry by strong sphalerons by suggesting that the singlet-doublet sector may be extended to allow for decays into Standard Model leptons, but it is important to perform a full calculation of the baryon asymmetry in the minimal singlet-doublet model including the strong sphaleron effects. 

 The singlet-doublet model has also been studied in the literature mostly in the context of dark matter \cite{Cohen:2011ec,Abe:2014gua,Basirnia:2016szw,Calibbi:2015nha}, but we point out that since in fermion induced EWBG the new fermions must couple strongly to the Higgs, the lightest singlet-doublet fermion would be ruled out as a dark matter candidate from direct detection experiments. Generically, if the relic density of singlet-doublet fermions is large, direct detection experiments severely constrain the model. One way to avoid potential limits from direct detection is to allow the lightest singlet-doublet neutral fermion to decay on cosmological timescales, a requirement that is easily fulfilled by allowing decays into Standard Model leptons as in section \ref{sec:BAS}. We leave the study of singlet-doublet relics and a detailed study of extensions of the model to accommodate the observed dark matter density for future work.
 
\section{Acknowledgments}
We thank David Shih and Patrick Meade for reading the manuscript and for helpful comments. We also thank Andrew Long, Carlos Wagner and Harikrishnan Ramani for useful discussions. We thank Eduardo Ponton for a cross check of the effective potential and Cristina Mondino for help with the \textsc{FeynRules} implementation. The work of D.E.U. while at Stony Brook University is supported by PHY-1620628. This work was initiated at the New High Energy Theory Center (NHETC), Rutgers University. The work of D.E.U. while at Rutgers was supported by DOE-SC0010008. 
\appendix

\section{Electroweak precision analysis}
\label{app:EWP}
The singlet-doublet fermion mass eigenstates couple to electroweak gauge bosons due to their doublet component and lead to corrections to the gauge boson self energies at one-loop. In this section we review the corresponding corrections and limits from electroweak precision observables. 

The one-loop corrections to the gauge boson self energies are most easily obtained in Dirac notation, so we define four component fermion fields
\begin{equation}
\Psi_i
\equiv
{\def\arraystretch{1}\tabcolsep=8pt
\left(
\begin{array}{c}
   \psi_i  \\ 
     \psi_i^\dagger \\ 
  \end{array}
       \right)
       }
       \qq
\Psi_+
\equiv
{\def\arraystretch{1}\tabcolsep=8pt
\left(
\begin{array}{c}
   \psi_+  \\ 
     \psi_-^\dagger \\ 
  \end{array}
       \right)
       }       
\end{equation}
where the three fields $\Psi_i$, $i=1,2,3$ are the four-component neutral singlet-doublet Majorana mass eigenstates of the mass matrix \eqref{eq:Mf} and $\Psi^+$ corresponds to a Dirac charged mass eigenstate. The corresponding mass terms in the Lagrangian are $m_i \bar{\Psi}_i \Psi_i - m_L \bar{\Psi}_+ \Psi_+$. The interactions of the four-component fermion fields with the photon, $Z$ and $W^\pm$ gauge bosons are
\begin{eqnarray}
\nonumber
&&
c^{\gamma} 
A_\mu 
\, 
\bar{\Psi}_+
\, \gamma^\mu 
\,
\Psi_+ 
+
c^{Z}_+
Z_\mu
 \,
\bar{\Psi}_+
\, \gamma^\mu 
\, \Psi_+ 
+
\frac{1}{2}
c^{Z}_{V_{ij}}
 Z_\mu
\,
 \bar{\Psi}_i
 \, \gamma^\mu 
 \, \Psi_j
+
\frac{1}{2}
c^{Z}_{A_{ij}}
 Z_\mu
 \bar{\Psi}_i
 \, \gamma^\mu \gamma_5
 \, \Psi_j
 \,
\\
\nonumber
&+&
\Big[
c^{W}_{V_i}
W^+_\mu
 \,
\bar{\Psi}_+
\, \gamma^\mu 
\,
 \Psi_i
+
c^{W}_{A_i}
W^+_\mu
 \,
\bar{\Psi}_+
\, \gamma^\mu \gamma_5 
\,
 \Psi_i
 +
\textrm{h.c}
\Big]\\
\end{eqnarray}
where
\begin{eqnarray}
\nonumber
c^{\gamma}
&=& 
e
\\
\nonumber
c^{Z}_+
&=&
\frac{1}{2}
\big(
\,
g_2 \cos \theta_W
-
g_1\sin \theta_W
\,
\big)
\\
\nonumber
c^{Z}_{V_{ij}}&=& 
\frac{1}{4}
\big(
\,
g_2 \cos \theta_W + g_1 \sin \theta_W
\,
\big)
\big(
\,
U_{3i}^*
U_{3j}
-
U_{2i}^*
U_{2j}
-\textrm{c.c.}
\,
\big)
\\
\nonumber
c^{Z}_{A_{ij}}&=& 
-
\frac{1}{4}
\big(
\,
g_2 \cos \theta_W + g_1 \sin \theta_W
\,
\big)
\big(
\,
U_{3i}^*
U_{3j}
-
U_{2i}^*
U_{2j}
+\textrm{c.c.}
\,
\big)
\\
\nonumber
c^{W}_{V_i}
&=&
\frac{g_2}{\sqrt{2}}
\big(
\,
U_{3i}
-
U_{2i}^*
\,
\big)
\\
c^{W}_{A_i}
&=&
-\frac{g_2}{\sqrt{2}}
\big(
\,
U_{3i}
+
U_{2i}^*
\,
\big)
\label{eq:gaugecouplings}
\end{eqnarray}
The W boson couplings $c^{W}_{V_i}$ and $c^{W}_{A_i}$ in \eqref{eq:gaugecouplings}, are not invariant under the discrete reparametrization and background symmetry transformations described in section \ref{sec:model}. However, all the one-loop corrections to the gauge boson self energies are invariants under both the reparametrization and background symmetry transformations. They are given by
\begin{eqnarray}
\nonumber
\Pi_{\gamma\gamma}(q^2)
&=&
c_\gamma^2
\,
\Pi_V(m_L,m_L,q^2)
\\
\Pi_{ZZ}(q^2)
&=&
\big(
c^Z_+
\big)^2
\,
\Pi_V(m_L,m_L,q^2)
\nonumber 
\\
\nonumber
&+&
\frac{1}{2}
c_{V_{ij}}^{Z*}
c_{V_{ij}}^Z
\,\,
\Pi_V(-m_i,-m_j,q^2)
+
\frac{1}{2}
c_{A_{ij}}^{Z*}
c_{A_{ij}}^Z
\,\,
\Pi_A(-m_i,-m_j,q^2)
\\
\Pi_{WW}(q^2)
&=&
c^{W *}_{V_i}
c^W_{V_i}
\,\,
\Pi_V(m_L,-m_i,q^2)
+
c^{W *}_{A_i}
c^W_{A_i}
\,\,
\Pi_A(m_L,-m_i,q^2)
\label{eq:selfenergies}
\end{eqnarray}
where the minus signs come from the different sign conventions for the corresponding Lagrangian and propagator masses for the fermions in the loops and the factors of $1/2$ correspond to symmetry factors in loops of Majorana fermions. The functions $\Pi_{V,A}$ are given in appendix \ref{app:selfenergies}.

The corrections to electroweak precision observables coming from new physics much heavier than the electroweak scale may be studied using the S,T,U parameter formalism \cite{Peskin:1990zt,Peskin:1991sw}. When the new physics is at or below the electroweak scale three additional V,W,X parameters are needed for the electroweak precision analysis  \cite{Maksymyk:1993zm}. For generality, in this work we also consider the case in which at least one of the singlet-doublet fermions is at or below the electroweak scale, so we make use of the full STUVWX analysis. We make use of the self-energies \eqref{eq:selfenergies} to calculate the STUVWX parameters defined in \cite{Maksymyk:1993zm}, from which we determine fourteen electroweak precision observables \cite{Batell:2013bka}
\begin{eqnarray}
\Gamma_Z
&=& 
 2.4950 - 0.0092 S + 0.026 T + 0.019 V - 0.020 X
 \nonumber
 \\
\sigma^0_{had}
&=& 
  41.484 + 0.014 S - 0.0098 T + 0.031 X
   \nonumber
 \\
R_\ell
&=& 20.743 - 0.062S + 0.042T - 0.14X
\nonumber \\
A_{FB}^\ell
&=&
  0.01626 - 0.0061S + 0.0042T - 0.013X\nonumber \\
A_\ell
&=&
0.1472 - 0.028S + 0.019T - 0.061X
\nonumber \\
A_c
&=&
0.6680 - 0.012S + 0.0084T - 0.027X
\nonumber \\
A_b
&=&
0.93463 - 0.0023S + 0.0016T - 0.0050X
\nonumber \\
A_{FB}^c
&=&
0.0738 - 0.015S + 0.010T - 0.033X
\nonumber \\
A_{FB}^b
&=&
0.1032 - 0.020S + 0.014T - 0.043X
\nonumber \\
R_c
&=&
  0.17226 - 0.00021S + 0.00015T - 0.00046X
  \nonumber \\
R_b
&=&
  0.21578 + 0.00013S - 0.000091T + 0.00030X
  \nonumber \\
s_{\theta_{\textrm{eff}}}^2
&=&
  0.23150 + 0.0035S - 0.0024T + 0.0078X
  \nonumber \\
m_W
&=&
80.364 - 0.28S + 0.43T + 0.35U
\nonumber \\
\Gamma_W
&=&
  2.091 - 0.015S + 0.023T + 0.018U + 0.016 W
  \label{eq:listew}
\end{eqnarray}
The Standard model input values for the electroweak observables above, which correspond to their values for STUVWX equal to zero, are taken from \cite{Baak:2014ora}. To obtain $95\%$ exclusion regions for our singlet-doublet model, we perform a chi-squared analysis on the fourteen precision observables and rule out the parameter space leading to $\chi^2 > 23.68$. The errors on the electroweak precision observables needed for the $\chi^2$ fit are taken from \cite{Batell:2013bka}.

\begin{figure}[htbp]
\begin{center}
\includegraphics[width=8cm]{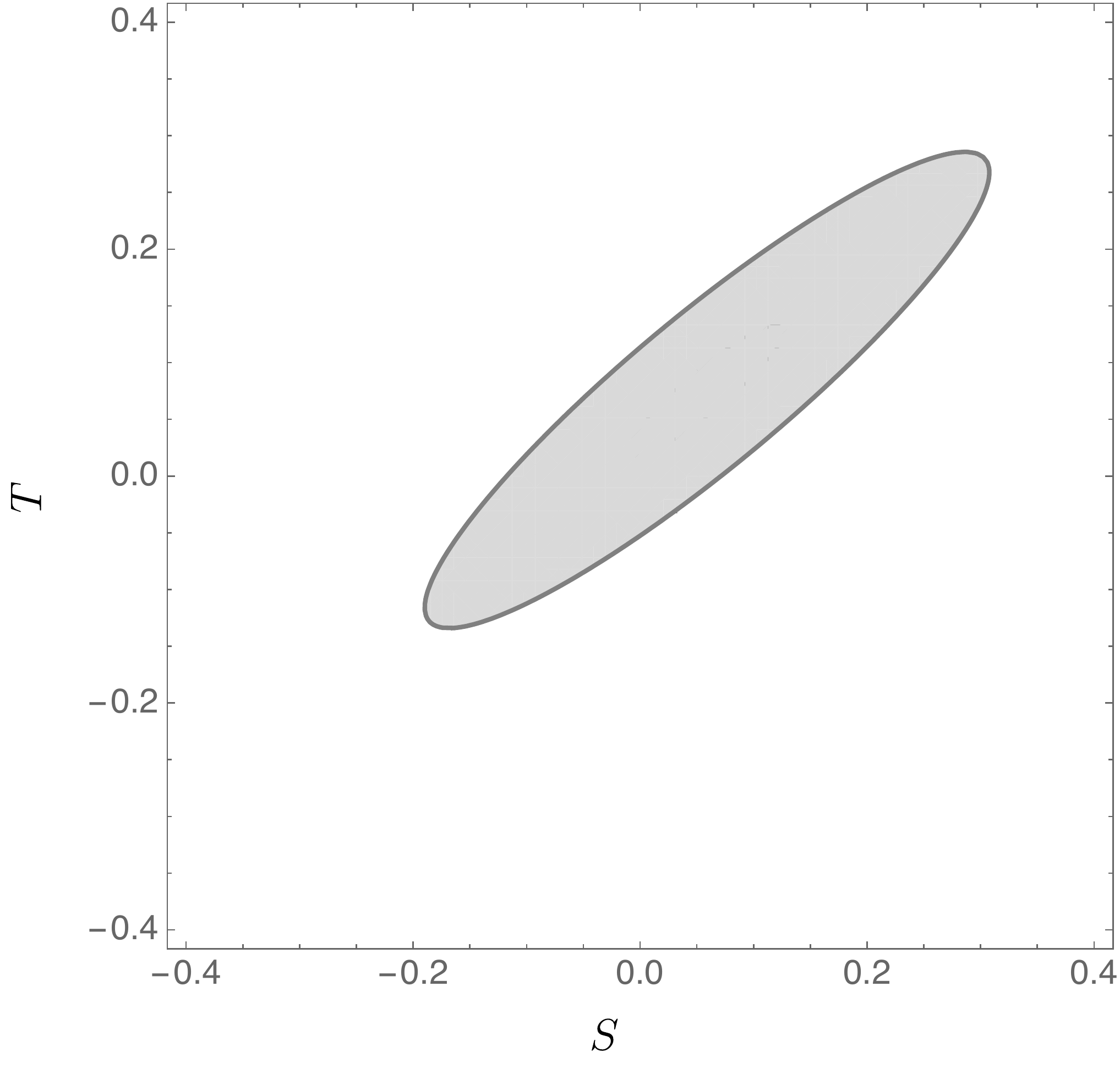}
\caption{Gray: allowed region in the S,T plane at $95\%$ confidence level, using a 14-dimensional $\chi^2 < 23.68$ test on the parameters \eqref{eq:listew}. In this figure,  we set the $U,V,W,X$ parameters to zero. Standard model experimental inputs and errors are taken from \cite{Baak:2014ora}.}
\label{fig:STfit}
\end{center}
\end{figure}

As a validation of our fit, in figure \ref{fig:STfit} we show in gray the $95\%$ confidence level allowed region in the $S,T$ plane, by setting $U,V,W,X$ to zero. The boundaries of the excluded region are within $\sim 10\%$ of the ones presented in reference \cite{Baak:2014ora}, which are obtained performing a similar fit.

\section{One-loop self-energies}
\label{app:selfenergies}
In this appendix we give explicit, analytic expressions for the one-loop self energies required for the electroweak precision analysis of appendix \eqref{app:EWP}. The self energies are calculated with \textit{Package-X} \cite{Patel:2015tea,Patel:2016fam}, in a field basis where the masses are real \footnote{\textit{Package-X} by default works only with positive masses. Here we allow the masses to be positive or negative. We thank Hiren Patel for the corresponding generalized expressions.}. The self energies are given by
\begin{eqnarray}
\nonumber
\Pi_{A \atop V}(m_1,m_2,q^2)
&=&
\frac{1}{24\pi^2}
\Bigg[
\,\,
\frac{1}{3q^2}
    \big(3 
    (m_1^2 
    -m_2^2)^2
    + 6 (m_1^2+m_2^2) q^2 
    \pm 36 m_1 m_2 q^2 
    - 10 q^4 
    \\
    \nonumber
    &&
    - 
     9 \gamma_Eq^2 
    (
    m_1
    \pm 
      m_2  
        )^2
        + 6 \gamma_E q^4
    \big) 
    \nonumber 
    \\
    &&
    - 
    ~
    \frac{1}{2  q^4}
    (m_1^6 
    - 3 m_1^4 m_2^2 
    + 3 m_1^2 m_2^4 
    - m_2^6 
    \pm 6 m_1^3 m_2 q^2 
    \mp 
    6 m_1 m_2^3 q^2 
    \nonumber
    \\
    &&
    -
    ~ 3 m_1^2 q^4 
    \mp 6 m_1 m_2 q^4 
    - 3 m_2^2 q^4 
    + 
    2 q^6) 
    \log
    \bigg(
    \frac{m_1^2}{m_2^2}
    \bigg)
      \nonumber
    \\
    &&
     +
     ~
     \frac{\Delta }{2 q^4}
   \big[
   (m_1^2-m_2^2)^2
   + (m_1^2+m_2^2) q^2 
   \pm 6 m_1 m_2 q^2 
    - 2 q^4
    \big]
       \nonumber
    \\
    && 
    \qq
    \log\Bigg(
    \frac{m_1^2 
    + m_2^2 
    - q^2 
    + 
  \Delta
  }
     { m_1^2 
     + m_2^2 
     - q^2 
     - 
     \Delta
     }
    \Bigg)
      \nonumber
    \\
    && 
    + 
    ~
  \big(3
  ( m_1
  \pm 
  m_2
  )^2
  - 2 q^2
  \big) 
  \Bigg[
  \,
  \frac{1}{\epsilon}
  + 
   \log
   \bigg(
   \frac{\mu^2}{m_1^2}
   \bigg)
   \,
   \Bigg]
   \,\,
   \Bigg]
\end{eqnarray}
where the upper signs corresponds to $\Pi_A$ and the lower signs to $\Pi_V$ and
\begin{equation}
\Delta
=
 \sqrt{
 (m_1^2-m_2^2)^2
  - 2 m_1^2 q^2 
  - 2 m_2^2 q^2 
  + 
   q^4}
\end{equation}

\section{$\beta$ functions in the singlet-doublet model}
\label{app:betafunctions}
In this appendix we give the beta function for the singlet-doublet Yukawas. We keep only terms proportional to singlet-doublet Yukawas, top quark Yukawa and gauge couplings $g_2,g_3$. The beta functions are given by
\begin{eqnarray}
\nonumber
\beta_{\lambda_{u}}\equiv \frac{d \lambda_u}{d \log \mu}
&=&
\frac{1}{16\pi^2}
\lambda_u
\,
\bigg(
\,
\frac{5}{2}
\lambda_u^* \lambda_u
+
3y_t^* y_t
-
\frac{9}{4}
g_2^2
\,
\bigg)
\\
\beta_{\lambda_{d}}
&=&
\frac{1}{16\pi^2}
\lambda_d
\,
\bigg(
\,
\frac{5}{2}
\lambda_d^* \lambda_d
+
3y_t^* y_t
-
\frac{9}{4}
g_2^2
\,
\bigg)
\nonumber 
\\
\beta_{y_{t}}
&=&
\frac{1}{16\pi^2}
y_t
\,
\bigg(
\,
\lambda_u^* \lambda_u
+
\lambda_d^* \lambda_d
+
\frac{9}{2}y_t^* y_t
-
8
g_3^2
-
\frac{9}{4}
g_2^2
\,
\bigg)
\end{eqnarray}

\bibliography{treeA_bib}

\end{document}